\documentclass[12pt]{elsart}
\usepackage{epsf}
\makeatletter
\DeclareRobustCommand*\cal{\@fontswitch\relax\mathcal}
\makeatother

\newcommand{\tauj}{\theta^{\ }_{\rm j}}
\newcommand{\taui}{\theta^{\ }_{\rm i}}
\newcommand{\ft}{\tilde{f}}
\newcommand{\gti}{\tilde{g}}
\newcommand{\be}{\begin{eqnarray}}
\newcommand{\ee}{\end{eqnarray}}

\newcommand{\la}{\langle}
\newcommand{\ra}{\rangle}
\begin{document}

\begin{frontmatter}

\title{Au+Au central collisions at 150, 250 and 400 AMeV energies
       in QMD with relativistic forces}
\author[Budapest]{J. N\'emeth},
\author[Budapest,GSI,HEID]{G. Papp} and
\author[GSI]{H. Feldmeier}
\address[Budapest]{Institute for Theoretical Physics, E\"otv\"os 
        Univ., H-1088 Budapest, Hungary,}
\address[GSI]{Gesellschaft f\"ur Schwerionenforschung, D-64220,
       Darmstadt, Germany,}
\address[HEID]{ITP, Univ. Heidelberg,
       Philosophenweg 19, D-69120 Germany.}

\date{\today}

\begin{abstract}
{Using the small acceleration approximation we derive a
relativistic scalar-vector force from a modified Zim\'anyi-Moszkowski
lagrangian based on $\sigma$, $\omega$ and $\rho$ meson exchanges. 
The momentum dependence of the force is fixed automatically by the
theory. We present an application of such a force in a QMD calculation
at intermediate energies comparing the results with the experimental
ones published recently by the FOPI collaboration. For most of the
quantities (number of intermediate mass fragments, ERAT, integrated
side flow, central flow, charge distributions, etc.)
we find agreement with the experimental
results.}\\[4mm] 
\end{abstract}

\end{frontmatter}


\section{Introduction}
\label{sec0}

The new experimental facilities~\cite{experiment} are able measure the
momenta of the protons and clusters and their charge which are created
in heavy ion collisions at intermediate energies (100 AMeV -- 2 AGeV) in
an exclusive way. Event-by-event analysis allows to 
test molecular dynamical models with respect to the degree of
equilibration, flow and cluster formation. Consequently, the
exisiting models are forced to reproduce not only one-body observables,
which are mainly governed by the mean-field and the continuity equation
but also many-body correlations which are necessary to create
intermediate mass fragments (IMF) at low excitation energy.

Nowadays it is popular to describe heavy ion collisions
with the help of transport models like
BUU~\cite{BUU}, QMD~\cite{AICH,frank}, etc. In these models the
interaction is divided into two parts, a smooth mean-field and a
random collision term. The mean-field force is often treated 
non-relativistically and the elastic two-particle collision cross section
is determined from the free nucleon-nucleon scattering. However, at
beam energies above 500 AMeV a relativitic treatment is necessary. Since
it has been recognized, that at least up to 2 AGeV the mean-field is
still of importance, one should include it in a relativistically invariant
form. At these energies particle creation in a hadronic medium is the
important issue. In order to disentangle the effect of the mean-field from
the medium dependence of the cross-sections one should try to check 
the validity of the applied mean-field at lower energies, where particle
creation is negligible.

In a previous publication by the authors~\cite{USHIP} 
a QMD code was developed in which the mean-field dynamics was based on
a relativistic 
lagrangian of the Walecka type. A covariant many-body Lagrange function
for point-like particles with scalar and vector two-body potentials was
derived, which still had the saturation property of the
original Walecka-type lagrangian. We checked this code at low energies
(50 -- 200 AMeV) in O--Br collisions and found, as many authors before,
that the Walecka type of mean-field force is not satisfactory for
QMD calculations due to the large values of the scalar and vector
coupling. Instead of adding third and fourth powers of scalar field in
the lagrangian~\cite{boguta} we used a derivative coupling proposed first
by Zim\'anyi and 
Moszkowski (ZM)~\cite{zimmos}. The advantage of this lagrangian is that 
the coupling constant determined by fitting nuclear saturation density
and energy are much smaller than for a linear coupling of the scalar
field. In addition the resulting incompressibility is two times smaller
(see Table~\ref{tab-coup}).

In this paper we deduce a relativistically invariant formulation of the
mean-field forces starting from the ZM lagrangian and using the small
accelaration approximation. We use the derived forces to investigate
central gold on gold collisions at 150, 250 and 400 AMeV 
and compare our results with the FOPI experiments~\cite{FOPI}.
 We also determined the
freeze out density and time as the function of the energy. In the
following we use $\hbar=c=1$ units, except where they are indicated
explicitely.

The paper is organized as follows. In Section 2  the
relativistic equations of motions for the nucleons are derived
generalizing the 
results of the previous paper~\cite{USHIP}. Section 3 gives details of the
QMD calculations. In Section 4 we compare our results with the
experimental data and in Section 5 the freeze out time and density are
determined. 
In Section 6 the distribution of intermediate mass fragments and their
dependence on various parameters is analyzed. Finally we conclude in
Section 7.

\section{Relativistic equations of motion for nucleons}

In this section we are sketching the derivation of the relativistic
equations of motion 
for the 4-positions and 4-momenta of the 
nucleons~\cite{USHIP,Fel1,Fel2}.
We start from a field theoretical lagrangian 
which includes besides a scalar field $\phi(x)$ and a vector field
$A_\alpha(x)$ an iso-vector vector field $\vec{B}_\alpha(x)$,
\begin{eqnarray}\label{Lagr1}
{\cal L}(x) &  = & \bar{\psi}(x) \Big( \gamma^\alpha i\partial_\alpha 
		- m f\!\!\left(\phi(x)\right) \Big)
\psi(x) \nonumber \\
& - & g_\omega \bar{\psi}(x) \gamma^\alpha \psi (x)  A_\alpha (x) 
	-g_\rho \bar{\psi}(x) \gamma^\alpha \frac 12 \vec{\tau} \psi (x) 
		\vec{B}_\alpha(x) \nonumber \\
& - & \frac{1}{2} \phi(x) (\partial_\alpha \partial^\alpha + \mu^2_\sigma)
\phi (x) \nonumber \\
& + & \frac{1}{2} A^\alpha(x) (\partial_\beta \partial^\beta + \mu^2_\omega)
A_\alpha (x) 
  + \frac 12 \partial^\alpha A_\alpha(x) \partial^\beta A_\beta(x) \\
& + & \frac 12 \vec{B}^\alpha(x) (\partial_\beta \partial^\beta +
	\mu^2_\rho) \vec{B}_\alpha(x)
	+\frac 12 \partial^\alpha \vec{B}_\alpha(x) \partial^\beta 
	\vec{B}_\beta(x)
\ . \nonumber
\end{eqnarray}
In order to avoid the stiff equation of state which results from a
linear coupling to the scalar field we use the proposal of
Zim\'anyi and Moszkowski~\cite{zimmos} to couple the scalar field in a
non-linear fashion as
\be
  f(\phi) = \frac 1{1+\frac{g_\sigma}m \phi} \,.
\ee

The field equations are
\begin{equation}\label{Dirac}
\Big\{ \gamma_{\alpha} (i \partial_\alpha - g_\omega  A_\alpha(x)
	-g_\rho \frac 12\vec{\tau} \vec{B}_\alpha(x) ) -
	m f\!\!\left(\phi(x)\right)\Big\}\ \psi(x) = 0
\end{equation}
for the nucleons,
\begin{equation}\label{Wavephi}
(\partial^\beta \partial_\beta + \mu^2_\sigma)\ \phi(x) = 
	-m f^\prime\!\!\left(\phi(x)\right)\ 
	\bar{\psi}(x)\psi(x)
\end{equation}
for the scalar field with $f^\prime=df/d\phi$ and
\be
\label{WaveA}
(\partial^\beta \partial_\beta + \mu_\omega^2)\ A^\alpha(x) =
   \partial^\alpha \partial_\beta A^\beta(x)
  + g_\omega \,  \bar{\psi}(x) \gamma^\alpha \psi(x) \\
(\partial^\beta \partial_\beta + \mu_\rho^2)\ \vec{B}^\alpha(x) =
   \partial^\alpha \partial_\beta \vec{B}^\beta(x)
  + g_\rho \,  \bar{\psi}(x) \gamma^\alpha \frac 12\vec{\tau} \psi(x) 
	\nonumber
\ee
for the vector fields. 

In the mean-field approximation the source terms are replaced
by their expectation values
$\la\bar{\psi}(x)\psi(x)\ra$, 
$\la\bar{\psi}(x) \gamma^\alpha \psi(x)\ra$ and 
$\la\bar{\psi}(x) \gamma^\alpha \vec{\tau} \psi(x)\ra$,
so that the scalar and vector fields
become classical fields. 

In addition we do not allow isospin rotations but conserve
neutron and proton currents individually. This means for the isospin
components
\renewcommand{\arraystretch}{0.7}
\be
\langle\bar{\psi}(x)\gamma^\alpha\vec{\tau}\psi(x)\rangle =
	\left(\begin{array}{c}
	  j_p^\alpha(x)-j_n^\alpha(x) \cr
	   0 \cr 0 \end{array}\right)
\ee\renewcommand{\arraystretch}{1}
with
\be
  \partial_\alpha j_p^\alpha(x) = 0 \quad \mbox{and} \quad
  \partial_\alpha j_n^\alpha(x) = 0  \,.
\ee
Here $j_p^\alpha(x)$ denotes the proton current and $j_n^\alpha(x)$ the
neutron current. Since the baryon current
\be
  \langle\bar{\psi}(x)\gamma^\alpha\psi(x)\rangle = j^\alpha(x) =
	j_p^\alpha(x)+j_n^\alpha(x)
\ee
is also conserved, the four-divergence of both classical fields, the
isoscalar 
$A^\alpha(x)$ and the isovector $\vec{B}^\alpha(x)$ vanish. Therefore
the terms $\partial^\alpha\partial_\beta A^\beta(x)$ and
$\partial^\alpha \partial_\beta \vec{B}^\beta(x)$ can be omitted in
Eq.~(\ref{WaveA}).

As we are using the relativistic scalar and vector fields only
for the mean-field part of the nuclear interaction we follow the
suggestion
\cite{Fel1,Fel2} to exclude radiation of mesonic fields 
from the very beginning by
using the action-at-a-distance formulation with the
symmetric Green's function of Wheeler and Feynman~\cite{Wheeler},
\begin{equation}
G(x-y) = \frac{1}{2}
 \Big( G^{advanced} (x-y)+  G^{retarded} (x-y)\Big) \,.
\end{equation}
Then the formal solutions of the Eq.~(\ref{Wavephi})
\begin{equation}\label{Phigreens}
\phi(x) = -m \int\! d^4 y \ G_\sigma (x-y) f^\prime\!\!\left(\phi(y)\right)
	\la\bar{\psi}(y)\psi(y)\ra
\label{gform}
\end{equation}
and of Eqs.~(\ref{WaveA})
\be
\label{Agreens}
A^\alpha(x) = g_\omega \int\! d^4 y \
 G_\omega (x-y) \la\bar{\psi}(y)\gamma^\alpha\psi(y)\ra \\
\vec{B}^\alpha(x) = \frac 12 g_\rho \int\! d^4 y \
 G_\rho (x-y) \la\bar{\psi}(y)\gamma^\alpha\vec{\tau}\psi(y)\ra \nonumber
\ee
fulfill the desired boundary condition of vanishing incoming and outgoing
free fields.

Eliminating the fields from the lagrangian (\ref{Lagr1}) leads to the
non-local action which contains only nucleon variables:
\begin{eqnarray}
\int d^4x\ {\cal L}(x)
	\hspace*{-4mm}&\hspace*{4mm}=\hspace*{-4mm}&\hspace*{4mm} 
\int d^4x\ \la\bar{\psi}(x) \gamma^\alpha i\partial_\alpha \psi(x)\ra
\nonumber \\
 &-& \int d^4x\ m \left[f\!\!\left(\phi(x)\right) - \frac 12 \phi(x)
	f^\prime\!\!\left(\phi(x)\right)\right] \la\bar{\psi}(x)\psi(x)\ra
\label{Action} \\
&-& \frac 12 g_\omega^2 \int d^4xd^4y\ 
	\la\bar{\psi}(x)\gamma^\alpha\psi(x)\ra\,
	G_\omega(x-y)\, 
	\la\bar{\psi}(y)\gamma_\alpha\psi(y)\ra \nonumber \\
&-& \frac 12 g_\rho^2 \int d^4xd^4y\ 
	\la\bar{\psi}(x)\gamma^\alpha\frac 12\vec{\tau}\psi(x)\ra
	\,G_\rho(x-y)\, 
	\la\bar{\psi}(y)\gamma_\alpha\frac 12\vec{\tau}\psi(y)\ra \nonumber
\end{eqnarray}
where $\phi(x)$ is given in terms of $\la\bar{\psi}(x)\psi(x)\ra$ by the
integral equation~(\ref{Phigreens}).
If quantum effects are negligible the scalar density and vector current density
can be represented in terms of the world lines of the nucleons as
\be
\label{rhos}
\rho_\sigma(x):= \la\bar{\psi}(x)\psi(x)\ra &\rightarrow&
 \sum_{\rm j=1}^{\rm A}
\int\! d \tauj \ \delta^4 (x-r_{\rm j}(\tauj))  \\
j^\mu(x):= \la\bar{\psi}(x)\gamma^\mu\psi(x)\ra &\rightarrow&
 \sum_{\rm j=1}^{\rm A}
\int\! d \tauj \ \delta^4 (x-r_{\rm j}(\tauj))\, 
u^\mu_{\rm j}(\tauj)\ ,
\label{rhos2} \\
\vec{i}^\mu(x):= \la\bar{\psi}(x)\gamma^\mu\vec{\tau}\psi(x)\ra 
	&\rightarrow& \sum_{\rm j=1}^{\rm A}
	\int\! d\tauj \ \delta^4 (x-r_{\rm j}(\tauj))\, 
	u^\mu_{\rm j}(\tauj) \vec{\tau}_{\rm j}\ ,
\label{rhos3}
\ee
where $r_{\rm j}(\tauj)$ denotes the world line of the nucleon 
$j$ at its proper time $\tauj$ and $u_{\rm j}(\tauj)$ its 
4-velocity. 
Furthermore $\vec{\tau}_{\rm j}$ denotes twice the isospin of the nucleon
$j$. Since there is no exchange of charged mesons in this lagrangian 
the isospin direction does not change in time so that 
 $\la{\rm neutron}|\vec{\tau}|{\rm neutron}\ra \equiv
 \vec{\tau}_{\rm neutron}=(-1,0,0)$ and
 $\la{\rm proton}|\vec{\tau}|{\rm proton}\ra \equiv
 \vec{\tau}_{\rm proton}=(1,0,0)$. This implies that only the first
component of the iso-vector current $\vec{i}^\mu(x)$ and the iso-vector
field $\vec{B}^\mu(x)$ contribute. Therefore we shall simply write
$i^\mu(x)$ and $B^\mu(x)$, respectively.

Nucleons in nuclei can however not be localized well enough for
treating them as classical particles on world lines. The Fermi momentum sets a
lower limit to the size of the nucleon wave packet in coordinate space
which is of 
the order of 2 fm. These effects, the Pauli principle and other quantum
effects, are properly taken care of in Antisymmetrized Molecular Dynamics
(AMD)~\cite{amd} and in Fermionic Molecular Dynamics (FMD)~\cite{fmd}.
However, presently neither the AMD nor the FMD equations of motion can be
solved numerically for large systems like gold on gold. Therefore, we mimic the
finite size effect of the wave packets by folding the finally obtained forces
with Gaussian density distributions for the nucleons. This "smearing" provides
also the prescription how to eliminate strong forces at short distances which
are taken care of by the random forces of the collision term.

Therefore, the forces to be derived in the following are only meant to
represent the long range part of the nuclear interactions which constitute a
mean field.
In this sense $r_{\rm j}(\tauj)$ is the mean world line of the nucleon and
$u_{\rm j}(\tauj)$ its mean 4-velocity.

Using the representation (\ref{rhos2})
the isoscalar field at a space-time point
$x$, for example, is given by integrals over past and future proper 
times $\tauj$ as
\begin{equation}\label{Phiworld}
B^\alpha(x) = g_\rho \sum_{\rm j=1}^{\rm A}
        \int\! d\tauj \ G_\rho(x-r_{\rm j}(\tauj))
	u_j^\alpha(\tauj) \frac 12\tau_{\rm j}
        \ .
\end{equation}

By identifying the different contribution of the action~(\ref{Action})
with their
representation in terms of mean positions and mean momenta of the 
nucleons one obtains the non-instantaneous
action-at-a-distance~\cite{Wheeler}:
\begin{eqnarray}\label{Action1}
\int\! d^4x \ {\cal L}(x) \hspace*{-8mm}&
	\hspace*{8mm}\rightarrow\hspace*{-8mm}&\hspace*{8mm} \nonumber \\
{\cal A} &=&  
	-\frac{1}{2} \sum^{\rm A}_{\rm i=1} \int\! d\taui
	 m_{\rm i}^*(\taui) u_{\rm i}(\taui)^2   
	  \nonumber \\
&-& m \sum^{\rm A}_{\rm i=1} \int\! d\taui\ 
	\left[f\!\!\left(\phi({\rm r_i}(\taui))\right) - \frac 12
	  \phi\left({\rm r_i}(\taui)\right) 
		f^\prime\!\!\left(\phi({\rm r_i}(\taui))\right) \right] \\
&-&\frac 12 g^2_\omega 
	\sum^{\rm A}_{\rm i,j=1}\!\!{}^{{}^{\mbox{\large $\prime$}}}
\int\! d\taui d\tauj \ G_\omega(r_{\rm i}(\taui) - 
  r_{\rm j}(\tauj))\ 
    u_{{\rm i}\alpha}(\taui) u_{\rm j}^\alpha(\tauj) \nonumber \\
&-&\frac 12 g^2_\rho 
\sum^{\rm A}_{\rm i,j=1}\!\!{}^{{}^{\mbox{\large $\prime$}}}
\int\! d\taui d\tauj \ G_\rho(r_{\rm i}(\taui) - 
	r_{\rm j}(\tauj))\ 
	u_{{\rm i}\alpha}(\taui) u_{\rm j}^\alpha(\tauj) 
	\frac 12\vec{\tau}_{\rm i} \frac 12\vec{\tau}_{\rm j} \nonumber \ ,
\end{eqnarray}
where the prime on the sum means ${\rm i}\neq{\rm j}$.

In this non-local Lagrange function the world lines $r_{\rm i}(\taui)$
of the particles are the variables. But only three components are
independent because the proper time $\theta$ relates the time to the
space component, or the square of the four-velocity equals one
($(dr/d\theta)^2=u^2=1$). Since we want manifest covariant equations of
motion Lagrange multipliers $m_{\rm i}^*(\taui)$ are introduced
in~(\ref{Action1}) and all four components of $r_{\rm i}$ are varied.

The result of the variation is
\be
\label{varAct}
  \delta {\cal A} \hspace*{-6mm}&\hspace*{6mm}=\hspace*{-6mm}&\hspace*{6mm}
	 -\sum^{\rm A}_{\rm i=1} \int\! d\taui
	m_{\rm i}^*(\taui) u^\alpha_{\rm i}(\taui) \frac d{d\taui} \
	  \delta r_{{\rm i}\alpha}(\taui) \nonumber \\
&-& \frac 12 m \sum^{\rm A}_{\rm i=1} \int\! d\taui
	\left[f^\prime\!(\phi_{\rm i})\ 
	  \delta \phi_{\rm i} - \phi_{\rm i}\ 
		\delta f^\prime\!(\phi_{\rm i}) \right] \\ 
&-& g^2_\omega
	\sum^{\rm A}_{\rm i,j=1}\!\!{}^{{}^{\mbox{\large $\prime$}}}
	 \int\!\!d\taui\! d\tauj\!
	\Big[ \partial^\alpha G_\omega^{\rm ij}\ 
	  \delta r_{{\rm i}\alpha}(\taui)\ 
	u_{{\rm i}\beta}(\taui) u^\beta_{\rm j}(\tauj) 
	+G_\omega^{\rm ij} 
	\left(\frac d{d\taui} \delta r_{{\rm i}\alpha}(\taui)\right)
	u^\alpha_{\rm j}(\tauj) \Big] \nonumber \\
&-& g^2_\rho
	\sum^{\rm A}_{\rm i,j=1}\!\!{}^{{}^{\mbox{\large $\prime$}}} 
	 \int\!\!d\taui\! d\tauj\!
	\Big[ \partial^\alpha G_\rho^{\rm ij}\  
	  \delta r_{{\rm i}\alpha}(\taui)\ 
	u_{{\rm i}\beta}(\taui) u^\beta_{\rm j}(\tauj) 
	+G_\rho^{\rm ij} 
	\left(\frac d{d\taui} \delta r_{{\rm i}\alpha}(\taui)\right)
	u^\alpha_{\rm j}(\tauj) \Big] \frac 12\vec{\tau}_{\rm i} 
	\frac 12\vec{\tau}_{\rm j}
	\nonumber
\ee
where $G^{\rm ij}\equiv G(r_{\rm i}(\taui)-r_{\rm j}(\tauj))$ and
$\phi_{\rm i}\equiv\phi(r_{\rm i}(\taui))$.

The variaton of the part containing the scalar field will be treated
below in more detail, as it is not straightforward due to the non-linear
coupling and the implicit integral equation~(\ref{Phigreens}). According
to~(\ref{gform}) and~(\ref{rhos}) the scalar field at 4-position 
 $r_{\rm i}$ is given by
\be
  \phi_{\rm i} \equiv \phi(r_{\rm i}) = 
	-m \sum^{\rm A}_{\rm i=1}\!{}^{{}^{\mbox{\large $\prime$}}}
	\int\!d\tauj\ G_\sigma^{\rm ij}\,f^\prime(\phi_{\rm j})
\ee
and the variation
\be
  \delta\phi_{\rm i}&=&-m 
	\sum^{\rm A}_{\rm j=1}\!{}^{{}^{\mbox{\large $\prime$}}}
	\int\! d\tauj
	\Big[ \delta G_\sigma^{\rm ij} f^\prime\!(\phi_{\rm j})+
	G_\sigma^{\rm ij}\ \delta
	f^\prime(\phi_{\rm j}) \Big] \,.
\ee
Inserting this into the second line of Eq.~(\ref{varAct}) yields
\be
  \delta{\cal A}_\phi &=& \frac {m^2}2 
	\sum^{\rm A}_{\rm i,j=1}\!\!{}^{{}^{\mbox{\large $\prime$}}} 
	\int\! d\taui d\tauj
	f^\prime\!(\phi_{\rm i})\ 
	  \delta G_\sigma^{\rm ij}\ f^\prime\!(\phi_{\rm j})
	\\
&+&  \frac {m^2}2 
	\sum^{\rm A}_{\rm i,j=1}\!\!{}^{{}^{\mbox{\large $\prime$}}} 
	\int\!\!d\taui\! d\tauj
  \Big[f^\prime\!(\phi_{\rm i})\ 
	G_\sigma^{\rm ij}\ \delta f^\prime\!(\phi_{\rm j})
 -\delta f^\prime\!(\phi_{\rm i})\ G_\sigma^{\rm ij}\ f^\prime\!(\phi_{\rm j})
	\Big] \,. \nonumber
\ee

Since $G_\sigma^{\rm ij}=G_\sigma^{\rm ji}$ is symmetric the last line
vanishes and 
\be
  \delta G^{\rm ij}=\partial^\alpha G(r_{\rm i}-r_{\rm j})
	\Big(\delta r_{{\rm i}\alpha}-\delta r_{{\rm j}\alpha}\Big) \,.
\ee

Integrating those terms which contain proper time derivatives
$d \delta r/d\theta$ by parts and requiring that $\delta {\cal A}=0$ for
arbitrary $\delta r_{\rm i}(\taui)$ finally yields
\be
\label{vartim}
\frac d{d\theta} \Big(m_{\rm i}^*(\theta) u_{\rm i}^\alpha\Big) &=&
	m\ f^\prime\!(\phi_{\rm i})\ \partial^\alpha \phi_{\rm i}
	+g_\omega \Big(\partial^\alpha A^\beta(r_{\rm i})
	-\partial^\beta A^\alpha(r_{\rm i}) \Big) u_{{\rm i}\beta} \\
  &+& g_\rho \tau_{\rm i} 
	\Big(\partial^\alpha B^\beta(r_{\rm i})
	-\partial^\beta B^\alpha(r_{\rm i}) \Big) u_{{\rm i}\beta}\,,
	\nonumber
\ee
where the fields are given according to 
Eqs.~(\ref{Phigreens})\ --~(\ref{Action}) by
\be
\label{eqphi}
  \phi(r_{\rm i}) &=& 
	-m\ \sum^{\rm A}_{\rm j\neq i} \int\! d\tauj
	G_\sigma^{\rm ij} f^\prime\!(\phi(r_{\rm j})) \,, \\
\label{eqA}
  A^\beta(r_{\rm i}) &=& 
	g_\omega \sum^{\rm A}_{\rm j\neq i} \int\! d\tauj
	G_\omega^{\rm ij}\ u^\beta_{\rm j} \,, \\
  B^\beta(r_{\rm i}) &=&
	g_\rho \sum^{\rm A}_{\rm j\neq i} \int\! d\tauj
	G_\rho^{\rm ij}\ u^\beta_{\rm j} \frac 12\tau_{\rm j}
\ee
and
\be
	u_{\rm i}(\theta) = \frac d{d\theta} r_{\rm i}(\theta) \,.
\ee

The Lagrange parameters $m_{\rm i}^*(\theta)$ are determined by
multiplying the above equation with $u_{{\rm i}\alpha}(\theta)$ and
using $d u^2_{\rm i}/d\theta=0$,
\be
  \frac{dm^*_{\rm i}(\theta)}{d\theta} &=&
	m f^\prime\!(\phi(r_{\rm i}))\ \partial^\alpha\phi(r_{\rm i})
	  u_{{\rm i}\alpha} \nonumber \\
&=& m f^\prime\!(\phi(r_{\rm i}))\ \frac d{d\theta} \phi(r_{\rm i})
	= m\ \frac d{d\theta} f\!(\phi(r_{\rm i})) \, ,
\ee
or
\be
  m^*_{\rm i}(\theta) = m f(\phi(r_{\rm i})) =
	\frac m{1+\frac{g_\sigma}m \phi(r_{\rm i}(\theta))} \,,
\label{eq:mstar}
\ee
where the integration constant has been set to zero so that for
$g_\sigma=0$, $m^*_{\rm i}=m$. Each particle has an effective mass
$m^*_{\rm i}$ which depends through the scalar field 
$\phi(r_{\rm i}(\theta))$ on all other particles in the same way as in
the original field-theory Lagrangian.

These derivations led us from a covariant Lagrangian for fields to the
following manifestly covariant equations of motion for worldlines,
\be
m f\!\Big(\phi(r_{\rm i})\Big)\ \frac{du^\alpha_{\rm i}}{d\theta} &=&
	m f^\prime\!(\phi(r_{\rm i}))\  
	\Big(\partial^\alpha\phi(r_{\rm i}) - 
	  \partial^\beta\phi(r_{\rm i})  u_{{\rm i}\beta}u^\alpha_{\rm i}
	\Big)  \\
&\phantom{a}& \hspace*{-20mm} +g_\omega 
	\Big(\partial^\alpha A^\beta(r_{\rm i}) - 
	  \partial^\beta A^\alpha(r_{\rm i})\Big)  u_{{\rm i}\beta}
	+ g_\rho \vec{\tau}_{\rm i}
	\Big(\partial^\alpha \vec{B}^\beta(r_{\rm i}) - 
	  \partial^\beta \vec{B}^\alpha(r_{\rm i})\Big)\  u_{{\rm i}\beta}
  \,.\nonumber
\ee
The coupling $m f^\prime\!(\phi)$ in front of the derivatives of
the scalar field always depends on the field itself (which can actually
also be choosen different from~(\ref{eq:mstar}), see~\cite{Lindner}),
\be
  m f^\prime(\phi) = -g_\sigma \left(\frac{m^*(\phi)}m\right)^2 \,,
\ee
the larger the field $\phi$ the weaker is the coupling.

The equations of motion which result from the Wheeler-Feynman action
extended by scalar fields which couple in a non-linear fashion
are completely equivalent to the classical field equations, provided
the system does not radiate. The reasoning is the same as given by
Wheeler and Feynman in~\cite{Wheeler} for the case of electrodynamics.

However, the Wheeler Feynman equations of motion cannot be 
solved in general 
as one needs to know the world lines for all
past and future times in order to calculate the fields which enter
the equations for the world lines~\cite{BelMartin}. Furthermore, the 
no-interaction theorem states that, except in the case of 
non-interacting particles, there exist no covariant
equations of motion for world lines in which only the 4-positions and
the 4-velocities at a given time enter, as it is the case in
non-relativistic mechanics.

In the following this no-interaction theorem is circumvented by
introducing an approximative solution to the non-local Wheeler Feynman
equations of motion. This is achieved by the so called
small acceleration approximation which does not assume small
velocities.

\subsection{Action-at-a-distance in the small acceleration 
approximation}

In order to conserve manifest covariance of the equations of motion
we introduce first the concept of a scalar time. For that the
whole Minkowski space is chronologically ordered by a set of 
space-like surfaces $S(x)$ which attribute to each 4-position
$x$ a scalar time $t$ by
\begin{equation}\label{Isochrone}
S(x)=t \ \ , \ \ \  \mbox{with} \ \ \ \partial^0 S(x) > 0 
 \ .
\end{equation}
The simplest choice for these isochrones, which we shall use 
in the following, is
$S(x)=\eta_\alpha x^a$, where $\eta_\alpha$ is a position
independent time-like 4-vector.

\begin{figure}[htbp]
\centerline{\epsfysize=8cm \epsfbox{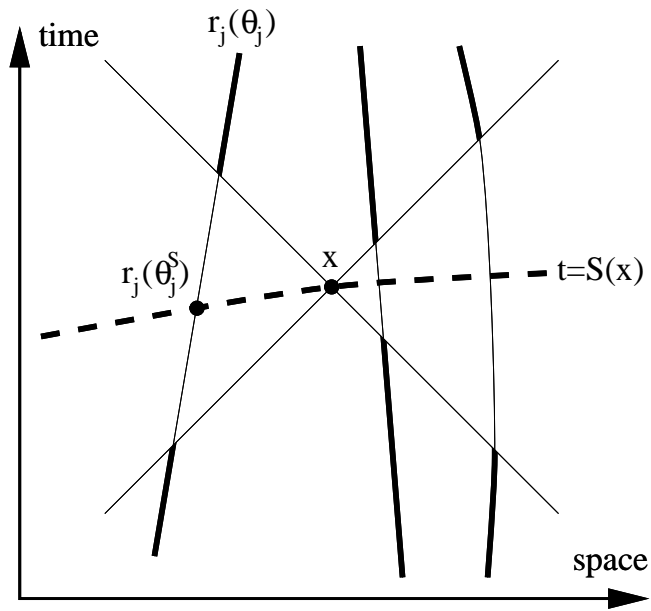}}
\caption{World lines and synchronizing hypersurface $S(x)$}
\label{fig-wline}
\end{figure}

Unlike in a collision the nucleons are not strongly accelerated by the
action of the mean-field. Therefore, in order to describe the motion
in the mean-field one may expand each world line
around the proper time $\theta^s_{\rm j}$ at which the particle is at 
a given isochrone, i.e. $t=S(r_{\rm j}(\theta^s_{\rm j}))$ for each $j$.
 
\begin{equation}\label{worldline}
r_{\rm j}^\alpha(\tauj)=r_{\rm j}^\alpha(\theta^s_{\rm j})
  +(\tauj-\theta^s_{\rm j})
u_{\rm j}^\alpha(\theta^s_{\rm j})+
\frac{1}{2}(\tauj-\theta^s_{\rm j})^2 \,
a_{\rm j}^\alpha(\theta^s_{\rm j})+\cdots \ .
\end{equation}
This way we are defining a synchronization prescription for all particles,
which does not
depend on the frame. For calculating the fields we neglect the 
quadratic term with the acceleration
$a_{\rm j}^\alpha(\theta^s_{\rm j})$ and all higher powers 
which leaves us with a straight world line in the vicinity of the
synchronizing time $t=S(r_{\rm j}(\theta^s_{\rm j}))$
(c.f. Fig.~\ref{fig-wline}).  
This is called ''small acceleration approximation".

The small acceleration approximation is of course best in the vicinity
of $r_{\rm j}^\alpha(\theta^s_{\rm j})$ and becomes worse further away.
As sketched in Fig.~\ref{fig-wline}, only those parts of the world lines
(thick lines), 
which are inside the light cone (centered at $x$), 
contribute to the field strength at point $x$. 
A world line which hits the light cone far away from $x$ may be badly
approximated by Eq. (\ref{worldline}), but for short range interactions 
a distant particle does not contribute anymore to the field at $x$.

Thus, the first condition for the validity of the approximation is
that the range $\mu^{-1}$ is small compared to the curvature of the 
world lines, i.e. the inverse of the acceleration.
The second condition is weak radiation, which is fulfilled when the
acceleration is small compared to the meson mass $\mu$. Both
conditions are actually the same, namely  
\begin{equation}\label{smallacc}
\mid a_\mu a^\mu \mid \ \ll \ \mu^2 .
\end{equation}

The assumption of small acceleration is justified if the $\phi$, 
$A^\alpha$ and $B^\alpha$ fields are only meant to be the mean-field
part of the 
nucleon-nucleon interaction in a hadronic surrounding. The hard collisions
between individual nucleons which are due to the repulsive core will 
cause large accelerations and also create new particles.
These hard collisions cannot be described within mean-field models of the
Walecka type.
Therefore, it is consistent to regard $\phi (x)$, $A^\alpha(x)$ and
$B^\alpha$ as Hartree
mean-fields which bring about only small accelerations and which 
are not radiated away from their sources.

Inserting the straight world line into Eq.
(\ref{Phiworld})
results in the easily understood situation that the field at a point $x$ is
just the sum of Lorentz-boosted Yukawa potentials which are traveling
along with the source:
\begin{equation}\label{Byuk}
B^\alpha(x)  =  \frac{g_\rho}{4\pi}  \sum_{\rm j} 
\frac{\exp\left\{-\mu_\rho \sqrt{R_{\rm j}(x)^2}\right\} }
{\sqrt{R_{\rm j}(x)^2}}\ u^\alpha_j(\theta^s_j) \frac 12\tau_j \ ,
\end{equation}
where $R_{\rm j}(x)^2$ is given by
\begin{equation}
R_{\rm j}(x)^2 = - (x-r_{\rm j}(\theta^s_{\rm j}))^2 +
\left[(x^\alpha-r_{\rm j}^\alpha(\theta^s_{\rm j}))\, 
u_{{\rm j}\alpha}(\theta^s_{\rm j})\right]^2 \ .
\end{equation}
The vector field is derived in an analogue fashion as
\begin{equation}\label{Ayuk}
A^\alpha(x) = \frac{g_\omega}{4\pi} \sum_{\rm j} 
\frac{\exp\left\{-\mu_\omega\sqrt{R_{\rm j}(x)^2}\right\} }
{\sqrt{R_{\rm j}(x)^2}}\, u_{\rm j}^\alpha(\theta^s_{\rm j}) \ .
\end{equation}
The scalar field results from the implicit Eq.~(\ref{eqphi}) in which we
assume that the scalar field does not vary much along the part of the
world line which contributes, {\it i.e.} 
$\phi(r_{\rm j}(\tauj))\approx\phi(r_{\rm j}(\theta^s_{\rm j}))$ and
hence can be replaced by the field at the synchronizing time. The result
is
\be
	\phi(x)=\frac{g_\sigma}{4\pi} \sum_{\rm j}
	\frac{\exp\left\{-\mu_\sigma \sqrt{R_{\rm j}(x)^2}\right\} }
		{\sqrt{R_{\rm j}(x)^2}}\ 
	\frac 1{\Big[ 1+\frac{g_\sigma}m \phi(r_{\rm j}(\theta^s_{\rm j}))
		\Big]^2}  \,.
\label{eq:phi1}
\ee

At this level of the approximation the causality problem with the advanced
part of the Green function is not present because the retarded and the 
advanced fields are identically the same when they are
created by charges moving on straight world lines. 
Therefore one may regard the fields as retarded only.
In addition the unsolved problem of radiation reaction, where 
the radiation acts back on the world lines \cite{Jackson},
does not occur because there are no radiation fields any more.

\subsection{Instantaneous action-at-a-distance}

In the spirit of the small acceleration assumption discussed in the previous
section one can use the straight line expansion in the action 
(\ref{Action1}) and perform the integration over 
$\tauj$. This results in an instantaneous 
action-at-a-distance where the Lorentz-boosted Yukawa fields appear again
and there is only one time, the  scalar synchronizing
time $t$,
\be
   {\cal A} = \int\!dt\ {\cal L}\Big(r_{\rm i}(t),u_{\rm i}(t)\Big) \,.
\ee
The Lagrange function is now
\begin{eqnarray}
\label{eq:lang}
{\cal L}\Big(r_{\rm i}(t),u_{\rm i}(t)\Big) &=& 
 -\sum_{\rm i=1}^A \  {m_{\rm i}^* \over{ u_{\rm i}^0 }} 
 \\ \nonumber &&\hspace*{-23mm} - {1 \over 2} 
\sum^{\rm A}_{\rm i,j=1}\!\!{}^{{}^{\mbox{\large $\prime$}}}\ 
{g^2_\sigma \over{ 4\pi u_{\rm i}^0 }} 
{{\rm exp}\left\{ -\mu_\sigma \sqrt{R_{\rm ij}^2(t)} \right\} \over{
\sqrt{R_{\rm ij}^2(t)} }} 
	\frac 1{\Big[ 1\!+\!\frac{g_\sigma}m \phi(r_{\rm i}(t))
		\Big]^2}
	\frac 1{\Big[ 1\!+\!\frac{g_\sigma}m \phi(r_{\rm j}(t)) 
		\Big]^2}  \nonumber \\
&&\hspace*{-23mm}- {1 \over 2} 
	\sum^{\rm A}_{\rm i,j=1}\!\!{}^{{}^{\mbox{\large $\prime$}}}\ 
	{g^2_\omega \over{ 4\pi
	u_{\rm i}^0 }}
  {{\rm exp}\left\{ -\mu_\omega \sqrt{R_{\rm ij}^2(t)} \right\} \over{
  \sqrt{R_{\rm ij}^2(t)} }} u_{\rm i\alpha}(t) u_{\rm j}^{\alpha}(t) 
	\\ &&\hspace*{-23mm}
	- {1 \over 2} 
	\sum^{\rm A}_{\rm i,j=1}\!\!{}^{{}^{\mbox{\large $\prime$}}}\ 
	{g^2_\rho \over{ 4\pi
	u_{\rm i}^0 }}
  {{\rm exp}\left\{ -\mu_\rho \sqrt{R_{\rm ij}^2(t)} \right\} \over{
  \sqrt{R_{\rm ij}^2(t)} }} u_{\rm i\alpha}(t) u_{\rm j}^{\alpha}(t) 
  \frac 12\tau_{\rm i} \frac 12\tau_{\rm j}  \, . \nonumber
\end{eqnarray}
The four-positions $r_{\rm i}^\alpha(t)\equiv 
r_{\rm i}^\alpha(\taui(t))$ and
4-velocities $u_{\rm i}^\alpha(t) \equiv u_{\rm i}^\alpha(\taui(t))$
are to be taken at the same scalar synchronizing time $t$ and
\begin{equation}
R_{\rm ij}^2(t) := - (r_{\rm i}(t) - r_{\rm j}(t))^2 + 
\Big[(r_{{\rm i}\beta}(t) -
r_{{\rm j}\beta}(t)) u_{\rm j}^\beta (t)\Big]^2
\end{equation}

The small acceleration approximation together with the introduction of a
synchronizing hypersurface $S(x)$ leads to an equal time lagrangian which is
Lorentz-scalar and written in a manifestly covariant way. 

Giving up explicit covariance and choosing $\eta=(1,0,0,0)$ in a special
coordinate frame, the positions and velocities take the form
\begin{equation}\label{eq:rcm}
r_{\rm i}(\taui(t)) = \Big(t\, , \,\vec{r}_{\rm i}(t)\Big) 
\ \ \ \mbox{and} \ \ \ u_{\rm i}(\taui(t))
= \frac{1}{\sqrt{1-\vec{v}_{\rm i}^{\,2}(t)}} \Big( 1\, ,\, 
  \vec{v}_{\rm i}(t) \Big) \ .
\end{equation}
With that a Lagrange function ${\cal L}\Big(\vec{r}_{\rm i}(t), 
\vec{v}_{\rm i}(t)\Big)$
can be defined which depends only on the independent variables and one time.

The advantage of the instantaneous lagrangian is that one can
define easily the hamiltonian and the total momentum, which are then strictly
conserved by the equations of motion.

\subsection{Hamilton equations of motion}

In the following we want to express the total hamiltonian as a function of the
positions $\vec{r}_{\rm i}$ and the canonical momenta $\vec{p}_{\rm i}$. 
Using the Lagrange function given in Eq.~(\ref{eq:lang})
the canonical momenta and the energy are determined
as
\begin{eqnarray}
\vec{p}_{\rm i} &=& {\partial {\cal L} \over{ \partial \vec{v}_{\rm i} }} 
\quad , \\
E &=& \sum_{\rm i} {\partial {\cal L} \over{ \partial \vec{v}_{\rm i} }}
\vec{v}_{\rm i} - \cal{L} \quad .\nonumber
\end{eqnarray}
To be able to perform these derivatives and get an analytical expression
for the energy we expand the Lagrange function~(\ref{eq:lang}) 
up to
third order in $(g_\sigma/m) \phi_{\rm i}$. Since the $\phi$ field 
depends on the nucleon density, this approximation is better at low
densities, and its validity should be checked during the calculation. We
come back to this point later.

Performing these derivatives the momenta  $\vec{p}_{\rm i}$ and the
energy $E$ of the system turn out to be
\begin{eqnarray}
\label{eq:mom}
\vec{p}_{\rm i} &=& M_{\rm i}\vec{u}_{\rm i} 
- \frac 12
  \sum^{\rm A}_{\rm j\neq i}\ 
    \widehat{f}_{\rm ji} u_{\rm i}^0 \left[
  \vec{r}_{\rm ij} (\vec{r}_{\rm ij} \vec{u}_{\rm i})
  + \vec{u}_{\rm i} (\vec{r}_{\rm ij} 
  \vec{u}_{\rm i})^2 \right] N_{\rm ij}+\vec{G}_{\rm i}\,, \\
\label{eq:enu}
E &=& \sum _{\rm i=1}^{\rm A} u_{\rm i}^0  M_{\rm i} 
 - \frac 12
\sum^{\rm A}_{\rm i,j=1}\!\!{}^{{}^{\mbox{\large $\prime$}}}
  \widehat{f}_{\rm ji}(\vec{r}_{\rm ij} \vec{u}_{\rm i})^2 
  \left( u_{\rm i}^0 \right)^2\!N_{\rm ij} \\
 &+&
\sum^{\rm A}_{\rm i=1}
  \frac{(\vec{G}_{\rm i} \vec{u}_{\rm i})}{u_{\rm i}^0}
+ \frac 12
\sum^{\rm A}_{\rm i,j=1}\!\!{}^{{}^{\mbox{\large $\prime$}}}
	g_{\rm ij} (u_{{\rm i}\alpha} u_{\rm j}^{\alpha}) \frac 1{u_{\rm i}^0}
	\,, \nonumber
\end{eqnarray}
where
\begin{eqnarray}
M_{\rm i} &=& m - 
	\frac 12
\sum^{\rm A}_{\rm j\neq i}\ 
  f_{\rm ij} +\frac 1m
	\sum^{\rm A}_{\rm j,k=1}\!\!{}^{{}^{\mbox{\large $\prime$}}}\ 
	  f_{\rm ij}f_{\rm jk} \\
 &-& \frac 1{m^2}
   \sum^{\rm A}_{\rm j,k,l=1}\!\!{}^{{}^{\mbox{\large $\prime$}}}\ 
	\left[ 2 f_{\rm ij}f_{\rm jk}f_{\rm kl} 
	  + \frac 32 f_{\rm ij}f_{\rm jk}f_{\rm jl}
	  - \frac 12 f_{\rm ij}f_{\rm ik}f_{\rm il} \right] \,, \nonumber\\
N_{\rm ij} &=& \frac 1{u_{\rm j}^0} -\frac 2m
	\sum^{\rm A}_{\rm k=1}\!\!{}^{{}^{\mbox{\large $\prime$}}}\ 
	  \frac{f_{\rm jk}}{u_{\rm j}^0} -\frac 2m
	\sum^{\rm A}_{\rm k=1}\!\!{}^{{}^{\mbox{\large $\prime$}}}\ 
	  \frac{f_{\rm kj}}{u_{\rm k}^0} +\frac 2{m^2}
	\sum^{\rm A}_{\rm k,l=1}\!\!{}^{{}^{\mbox{\large $\prime$}}}\ \left[
	  \frac{2f_{\rm ik}f_{\rm kl}}{u_{\rm j}^0} +\right. \\
 && \left.
	  +\frac{2f_{\rm il}f_{\rm kj}}{u_{\rm k}^0} 
	  +\frac{2f_{\rm kl}f_{\rm jl}}{u_{\rm k}^0} 
	  +\frac{3f_{\rm ik}f_{\rm il}}{2 u_{\rm j}^0} 
	  +\frac{3f_{\rm kj}f_{\rm jl}}{u_{\rm k}^0} 
	  -\frac{3f_{\rm jk}f_{\rm jl}}{2u_{\rm j}^0} \right] \nonumber\\
\vec{G}_{\rm i} &=&
 {1\over 2} 
\sum^{\rm A}_{\rm j\neq i}\ 
  \left( g_{\rm ij} + {u_{\rm i}^0
\over{ u_{\rm j}^0 }} g_{\rm ji} \right) \vec{u}_{\rm j}
- {1\over 2} 
\sum^{\rm A}_{\rm j\neq i}\ 
  g_{\rm ji} \left[
\left( u_{\rm i}^0 \right)^2 - {u_{\rm i}^0 \over{ u_{\rm j}^0 }} 
  \vec{u}_{\rm i} \vec{u}_{\rm j} \right] \vec{u}_{\rm i} \\
&+& {1\over 2} 
\sum^{\rm A}_{\rm j\neq i}\
  \widehat{g}_{\rm ji} \left[
\left( u_{\rm i}^0 \right)^2 - {u_{\rm i}^0 \over{ u_{\rm j}^0 }} 
  \vec{u}_{\rm i} \vec{u}_{\rm j} \right] 
 \left[
  \vec{r}_{\rm ij} (\vec{r}_{\rm ij} \vec{u}_{\rm i}) +
  \vec{u}_{\rm i} (\vec{r}_{\rm ij} \vec{u}_{\rm i})^2
 \right] \, . \nonumber
\end{eqnarray}

In these expressions $f_{\rm ij}$ and $g_{\rm ij}$ are
\begin{eqnarray}
\label{eq:fdef1}
f_{\rm ij} \equiv f(R_{\rm ij}) = {g_\sigma^2\over{4 \pi}}\ {{\rm exp}\left\{
  -\mu_\sigma \rm R_{ij}\right\} \over{R_{\rm ij}}} \,&,& \quad
  R_{\rm ij}^2 = \vec{r}_{\rm ij}^2 + (\vec{r}_{\rm ij}
  \vec{u}_{\rm j})^2 \,, \quad 
  \vec{r}_{\rm ij} = \vec{r}_{\rm i} - \vec{r}_{\rm j} \, ,
  \nonumber \\
g_{\rm ij} = {g_\omega^2\over{4 \pi}}\ {{\rm exp}\left\{ -\mu_\omega
  \rm R_{ij}\right\} \over{R_{\rm ij}}} 
  &+& {g_\varrho^2\over{4 \pi}}\ {{\rm exp}\left\{ -\mu_\varrho
  \rm R_{ij}\right\} \over{R_{\rm ij}}}
	\frac 12 \tau_{\rm i} \frac 12 \tau_{\rm j}
  \, , \nonumber \\
\widehat{f}_{\rm ji} \equiv -{1 \over{ R_{\rm ji} }} {d \over{ d R_{\rm ji} }}
 f(R_{\rm ji}) =  {g_\sigma^2\over{4 \pi}}\ 
	&\hspace*{-5mm}\phantom{+}& \hspace*{-5mm}
  \ (1+\mu_\sigma R_{\rm ji}) 
 {{\rm exp}\left\{ -\mu_\sigma \rm R_{ji}\right\} \over{ R_{\rm ji}^3 }}\,,
\label{eq:fdef2}
\end{eqnarray}
and $\widehat{g}_{\rm ji}$ is defined in the same way as 
$\widehat{f}_{\rm ji}$.
It is worthwhile to note that $f_{\rm ij} \neq f_{\rm ji}$.

Let us first prove that for homogeneous and isotropic nuclear matter,
like in the original mean-field picture~\cite{serwal}
the vector potential will not contribute to the canonical momentum.
In that case the summation can be replaced by
an integration over space and momentum, folded with the phase space 
distribution 
function. Since for homogeneous nuclear matter the distribution function is
independent of the position, the space integral can be carried out
immediately and the remaining part is written as a summation over momenta.
Finally for infinite nuclear matter we get
\be
\vec{p}_{\rm i} &=& m^* \vec{u}_{\rm i} \,,\quad
	m^*=m\!-\!B\!+\!\frac{3B^2}m\!-\!12 \frac{B^3}{m^2} \,,\quad  
    B=\left( {g_\sigma\over{\mu_\sigma }} \right)^2
	  \frac 1{\Omega}\sum_{\rm i} \frac 1{u_{\rm i}^0}\\
 E_{nm} &=& \sum_{\rm i}\left[ m^* u_{\rm i}^0 +
	\frac 1{2u_{\rm i}^0} \left(B-4 \frac{B^2}m+18 \frac{B^3}{m^2}\right)
	+\frac 12\left( {g_\omega\over{\mu_\omega }} \right)^2
	  \left( j^0 \right)\right] \,.  \nonumber
\label{eq:pnm1}
\ee
Expression~(\ref{eq:pnm1}) is exactly the one in Ref.~\cite{zimmos}
for low density expansion.

It is worthwhile to mention that the small acceleration approximation which
introduces $R_{\rm ij}$ instead of $\mid \vec{r}_{\rm ij}\mid$ in 
Eq.~(\ref{eq:fdef1}), is
definitely needed to get back the relativistic mean-field result for nuclear 
matter. Hence we take for the strengths $g_\sigma$ and $g_\omega$ 
the values obtained from the saturation properties of the nuclear
matter~\cite{Lindner}.

To be able to apply the above method for QMD calculation we have to
get the energy as a function of $\vec{p}_{\rm i}$ and 
 $\vec{r}_{\rm i}$. In order to 
get this we use Eq.~(\ref{eq:mom}) to express $\vec{u}_{\rm i}$
as a function of the momentum $\vec{p}_{\rm i}$ by expanding up to third
order in $f_{\rm ij}/m$, $g_{\rm ij}/m$ and $\vec{p}_{\rm i}^2/m^2$.
After substituting $\vec{u}_{\rm i}$ into Eq.~(\ref{eq:enu}) the energy
is 
\begin{eqnarray}
{E\over{mc^2}} &=& 
	\underbrace{\rule[-10mm]{0pt}{10mm}
	\sum_{\rm i}\ \sqrt{1+\vec{\tilde{p}}_{\rm i}^2}
		}_{\mbox{\large $E_{kin}$}}
  \underbrace{- {1\over 2} 
  \sum^{\rm A}_{\rm i,j=1}\!\!{}^{{}^{\mbox{\large $\prime$}}}\ 
    \frac{\ft_{\rm ij}}%
	  {\sqrt{1+\vec{\tilde{p}}_{\rm i}^2}\left(1+2
    {\displaystyle \sum_{\rm k=1}\!\!{}^{{}^{\mbox{\large $\prime$}}}}\ 
	   f_{\rm jk} \right)} }_{\mbox{\large $E_S$}}
   \underbrace{\rule[-10mm]{0pt}{10mm} + \frac 12
  \sum^{\rm A}_{\rm i,j=1}\!\!{}^{{}^{\mbox{\large $\prime$}}}\ 
  \gti_{\rm ji} \sqrt{1+\vec{\tilde{p}}_{\rm i}^2} \nonumber
	}_{\mbox{\large $E_V$}} \\[2mm]
&-& \hspace*{-5mm}\underbrace{\hspace*{5mm}%
\label{eq:en}
	 \frac 12
  \sum^{\rm A}_{\rm i,j=1}\!\!{}^{{}^{\mbox{\large $\prime$}}}\ 
  \gti_{\rm ji} 
  {(\vec{\tilde{p}}_{\rm i} \vec{\tilde{p}}_{\rm j}) \over{
  \sqrt{1+\vec{\tilde{p}}_{\rm i}^2}}}}_{\mbox{\large $E_{p_ip_j}$}} 
	\\&+& \hspace*{-5mm}\underbrace{\hspace*{5mm} \frac 12
  \sum_{\rm i}\ \left( 
    \sum^{\rm A}_{\rm i,j=1}\!\!{}^{{}^{\mbox{\large $\prime$}}}\ 
    \ft_{\rm ji} \vec{\tilde{p}}_{\rm i}
  -\!\!\sum^{\rm A}_{\rm i,j=1}\!\!{}^{{}^{\mbox{\large $\prime$}}}\ 
  \gti_{\rm ji} \vec{\tilde{p}}_{\rm j} 
  \right)^2 
	\!+\!
  \sum^{\rm A}_{\rm i,j,k,l=1}\!\!\!\!{}^{{}^{\mbox{\large $\prime$}}}\ 
    \!\!\left( \ft_{\rm ij}\ft_{\rm jk}\ft_{\rm kl}
      \!-\!\ft_{\rm ij}\ft_{\rm ik}\ft_{\rm il}
     \!-\!\ft_{\rm ij}\ft_{\rm jk}\ft_{\rm kl} \right) }_{\mbox{\large $E_3$}} 
	\nonumber
\end{eqnarray}

\noindent where $\ft_{\rm ij}$ and $\gti_{\rm ij}$ are now functions of
\begin{eqnarray}
R_{\rm ij}^2 &=& \vec{r}_{\rm ij}^2 + ( \vec{r}_{\rm ij}
\vec{\tilde{p}}_{\rm j} )^2 \quad ,
\end{eqnarray}

\noindent with $\vec{\tilde{p}}_{\rm i} = \vec{p}_{\rm i}/mc$,
 $\ft_{\rm ij}=f_{\rm ij}/mc^2$, $\gti_{\rm ij}=g_{\rm ij}/mc^2$. 
$E_{kin}$, $E_S$, $E_V$, $E_{p_ip_j}$ denote the kinetic,
scalar, vector term, respectively and $E_3$ is of third order
and hence small, as we show later.

From Eq.~(\ref{eq:en}) the Hamilton equations 
\begin{equation}
{d \over{ dt }} \vec{r}_{\rm i} = 
{\partial E\over{\partial \vec{p}_{\rm i}}} \qquad 
{d \over{ dt }} \vec{p}_{\rm i} = -\ 
  {\partial E\over{\partial \vec{r}_{\rm i}}}
\label{eq:eom}
\end{equation}
are calculated easily.

\section{Details of the QMD calculations}

Quantum Molecular Dynamics is a classical many-body theory in which some
quantum features
due to the fermionic nature of nucleons are simulated.  For the details of
the theory we would like to refer the reader to the works of 
Aichelin~\cite{AICH} and the
Frankfurt group~\cite{frank}. 

\subsection{The initial conditions}

Although in our calculations up to now we considered the nucleons as point
particles, to determine the initial space and momentum distribution of the 
ground-state nuclei in their own rest frame we used the same procedure as 
described in Ref.~\cite{AICH} and~\cite{judqmd,paula}. The nucleons were
characterized both in coordinate and momentum space with gaussians, with
the Wigner density
\be
f_{n,p}(\vec{r},\vec{p},t) = \frac 1{(\pi\hbar)^3} \sum_i
	e^{-\alpha^2 |\vec{r}-\vec{r}_i(t)|^2} 
	e^{-\beta^2 |\vec{p}-\vec{p}_i(t)|^2} \,.
\label{e-packet}
\ee
The width $\alpha$ was chosen such that placing the nucleons on a
lattice for infinite nuclear matter calculation we get a
smooth distribution with proper binding energy and saturation
density. Since the scalar and vector interactions of the lagrangian
describe the long-range part of the nucleon-nucleon interaction, the
mean-field, whereas the short range repulsion is treated explicitely by
the collision term, a cutoff has to be introduced at short distances to
separate both. This is done by folding the Yukawa functions 
 $f_{\rm ij}=f(R_{\rm ij})$ with the gaussian density distribution
introduced in Eq.~(\ref{e-packet}) for the phase-space 
density~\cite{USHIP}. Due to the folding the forces depend on the width
parameter $\alpha$. A proper choice should give a reasonably smooth mean
field at nuclear matter density and at the same time be in accord with
the cross section which determines the distance of closest approach
below which random scattering occurs.
We found, that the best value of $\alpha$, satisfying the above
requirements was $\alpha\approx 0.55$fm$^{-1}$. 
Choosing for example $\alpha=0.7 {\rm fm}^{-1}$, we could
not get a satisfactorily smooth distribution for nuclear matter on the
lattice. The parameter $\beta$ in~(\ref{e-packet}) was determined as
 $\beta=1/(\hbar\alpha)$.  

We construct the ground state of the gold nucleus by
 distributing the centers of the Gaussians so that the resulting 
configurations yield a binding energy per nucleon within 0.1 MeV of the
experimental value of 7.9 MeV, and 
 $|\vec{r}_{\rm i}-\vec{r}_{\rm j}| |\vec{p}_{\rm i}-\vec{p}_{\rm j}| > d$,
where $d$ is in the order of $\hbar$.
The initialization of the colliding
system is performed by boosting two such nuclei
relativistically towards each other, so that in the center of momentum frame
two Lorentz contracted nuclei are colliding.

\subsection{The mean-field forces}

The ZM forces derived in the previous section give saturation without
the introduction of a nonlinear scalar potential, originally proposed by
Boguta~\cite{boguta}, and used in many mean-field
calculations. This way we have less free parameters, than
usual. 
Furthermore fixing the masses of the scalar and vector mesons to
their generally accepted values 
the only remaining free parameters 
are the
meson coupling constants, $g_\sigma$, $g_\omega$ and $g_\rho$, which are
determined from the ground state properties of nuclear matter and the
symmetry energy. We summarize the parameters used in this work and the
resulting properties of the ZM force together with those of Walecka
lagrangian in Table~\ref{tab-coup}.  As one can see, the ZM force needs much
smaller values of the coupling constants, and yields also a much smaller
compression coefficient. In our calculation we are using the expanded
form~(\ref{eq:en}) of the exact energy expression~(\ref{eq:enu}), and in
order to keep the nuclear matter parameters, we modified the ZM
parameters accordingly. The values of the parameters for different
forces are summarized in Table~\ref{tab-coup}. The Coulomb energy
although not explicitely mentioned, is included as a zero mass vector
interaction.
\begin{table}[t]
\begin{center}
\begin{tabular}{||l|c|c|c|c|c|c|c|c|} \hline
& $g_\sigma$ & $g_\omega$ & $g_{\varrho}$ & $\mu_\sigma c^2$ & 
$\mu_\omega  c^2\!=\!\mu_{\varrho} c^2$ &
$m^*$ & $\kappa$ & $\varrho_0$ \\ \hline \hline 
W & 11.04 & 13.74 & 7.0 & 550 & 783 & 0.54 & 550.8 & 0.148 \\ \hline 
ZM & 7.84 & 6.67 & 7.0 & 550 & 783 & 0.85 & 224.7 & 0.160 \\ \hline
used & 7.22 & 5.58 & 7.0 & 550 & 783 & 0.8 & 212.1 & 0.158\\ \hline
\end{tabular}
\end{center}
\caption{Parameters of the Walecka, Zim\'anyi-Moszkowski and our
forces. The meson masses are given in MeV/c$^2$, the
compressibility in units of MeV and the densities are in fm$^{-3}$.}
\label{tab-coup}
\end{table}

We would like to emphasize, that no additional momentum dependent forces are
introduced, the momentum dependence is a generic property of the
relativistic force and
cannot be changed. It is therefore interesting to compare the predicted
value of observables, which depend strongly on the momentum
dependence, like different flow values, with the measured ones.

\subsection{The treatment of the collision term}

For the cross section in the collision term we used the Cugnon
one~\cite{cug} and for the Pauli blocking the
same method as is described in Ref~\cite{USHIP,judqmd}.  The individual
two-body collision is calculated in a relativistically covariant way, as
given in~\cite{wolf}, the only difference is that instead of the bare
mass of the nucleons, $m$, we use the effective mass, $m^*$, as
explained below.

In an interacting many-body system single-particle energies 
$e_{\rm i}$ cannot be defined such that ($e_{\rm i},\vec{p}_{\rm i}$)
form a 4-vector.
Only the total momentum $\vec{P}$ and total energy $E$ combine to a 4-vector.
However, we still try to deduce a single particle energy, which we use for
the conservation of energy in each binary collision. 
Equation (\ref{eq:en}) in leading order in $\vec{\tilde{p}}_{\rm i}$
and $\ft_{\rm i}, \gti_{\rm i}$ can be written as
\begin{equation}
\label{eq:efm}
{E\over{mc^2}} = \sum^A_{\rm i=1}\ 
	\sqrt{1\!+\!\vec{\tilde{p}}_{\rm i}^2}-\!{1\over 2} 
\sum^A_{\stackrel{{\scriptstyle \rm i,j=1}}{\rm j \neq i}}\ 
	{\tilde{f}_{\rm ij} \over{ 1\!+\!2 {\displaystyle 
		\sum_{\stackrel{{\scriptstyle \rm k}}{\rm k \neq j}}\ 
		\tilde{f}_{\rm jk} } }}
		{1 \over{ \sqrt{1\!+\!\vec{\tilde{p}}_{\rm i}^2}}} + {1\over 2} 
\sum^A_{\stackrel{{\scriptstyle \rm ij=1}}{\rm j \neq i}}\ 
\tilde{g}_{\rm ji} \sqrt{1\!+\!\vec{\tilde{p}}_{\rm i}^2} 
\end{equation}

There are two ways to define the effective mass $m^*_i$
appearing in the collisions:

Case A: If we neglect the momentum dependence in $\tilde{f}_{\rm ij}$ and
$\tilde{g}_{\rm ij}$, the energy of the i$^{\rm th}$ particle can be written
as
\begin{eqnarray}
\label{eq:sinen}
\frac{\epsilon_{\rm i}}{mc^2} = 
{E(A) - E_{\rm i}(A-1) \over{ mc^2 }} &=& \sqrt{ 1 +
\vec{\tilde{p}}_{\rm i}^2 } 
- {1 \over 2}
\sum_{\stackrel{{\scriptstyle \rm j}}{\rm j \neq i}}\ 
  {\tilde{f}_{\rm ij} \over {
\sqrt{ 1 + \vec{\tilde{p}}_{\rm i}^2 } }} 
{1 \over{ 1 + 2 {\displaystyle 
  \sum_{\stackrel{\scriptstyle \rm k}{\rm k \neq j}}\ 
    \tilde{f}_{\rm jk} } }} \\
&& 
    +{1 \over 2} 
\sum_{\stackrel{{\scriptstyle \rm j}}{\rm j \neq i}}\ 
  \tilde{g}_{\rm ji} \sqrt{ 1 + \vec{\tilde{p}}_{\rm i}^2 } 
+ \mbox{terms independent of }\  \vec{\tilde{p}}_{\rm i} \ ,\nonumber
\end{eqnarray}
where $E_{\rm i}(A-1)$ means the energy of the system when particle i is taken
out.

We can expand Eq.~(\ref{eq:sinen}) in $\vec{\tilde{p}}_{\rm i}^2$ and get
$$
\frac{\epsilon_{\rm i}}{mc^2} 
  = {1 \over 2} \vec{\tilde{p}}_{\rm i}^2 \left[ 1 + {1 \over 2}
\sum_{\stackrel{{\scriptstyle \rm j}}{\rm j \neq i}}\ 
  {\tilde{f}_{\rm ij} \over{ 1 + 2 {\displaystyle 
  \sum_{\stackrel{{\scriptstyle \rm k}}{\rm k \neq j}}\ 
\tilde{f}_{\rm jk} } }} + {1 \over 2} 
\sum_{\stackrel{{\scriptstyle \rm j}}{\rm j \neq i}}\ 
  \tilde{g}_{\rm ji} \right] + \epsilon_{\rm i0} \,.
$$
This can be written as
$$
\frac{\epsilon_{\rm i}}{mc^2} = 
	\sqrt{ \left( \frac{m^*_{\rm i}}m \right)^2 + 
\vec{\tilde{p}}_{\rm i}^2 } + \epsilon^{\prime}_{\rm i0} \quad ,
$$
where the effective mass $m^*_{\rm i}$ turns out to be
\begin{equation}
\label{eq:msr}
{m^*_{\rm i} \over m} = 1 
  - {1 \over 2} 
\sum_{\stackrel{{\scriptstyle \rm j}}{\rm j \neq i}}\ 
  {\tilde{f}_{\rm ij} \over{ 
    1 + 2 {\displaystyle 
  \sum_{\stackrel{{\scriptstyle \rm k}}{\rm k \neq j}}\ 
  \tilde{f}_{\rm jk} } }} 
  - {1 \over 2}
  \sum_{\stackrel{{\scriptstyle \rm j}}{\rm j \neq i}}\ 
   \tilde{g}_{\rm ji}
	\,,
\end{equation}
while $\epsilon_{\rm i0}$ and $\epsilon^{\prime}_{\rm i0}$ are
quantities independent of $\vec{\tilde{p}}_{\rm i}$.

Case B: If we assume, that the momentum dependence of 
$\tilde{f}_{\rm ij}$ and $\tilde{g}_{\rm ji}$ is the same as in nuclear
matter, that is
\begin{equation}
E = \sum_{\rm i}
    \sqrt{ 1\!+\!\vec{\tilde{p}}_{\rm i}^2 } 
  - {1\over 2} \sum_{\stackrel{{\scriptstyle \rm i,j}}{\rm j \neq i}}
    {\ft(r_{\rm ij}) \over{ \sqrt{ 1\!+\!\vec{\tilde{p}}_{\rm i}^2 } 
        \sqrt{ 1\!+\!\vec{\tilde{p}}_{\rm j}^2 } }}
  { 1 \over { 1\!+\!2 {\displaystyle 
    \sum_{\stackrel{{\scriptstyle \rm k}}{\rm k \neq j}}\ 
    \ft(r_{\rm jk}) } }}
  + {1\over 2} \sum_{\stackrel{{\scriptstyle \rm i,j}}{\rm j \neq i}}
    \gti(r_{\rm ji}) \,,
\end{equation}
and the effective mass for finite systems turns out
to be
\begin{equation}
\label{eq:msi}
{m^*_{\rm i} \over m} = 1 -
  \sum_{\stackrel{{\scriptstyle \rm j}}{\rm j \neq i}}\ 
  {\ft(R_{\rm ij}) \over{
    1 + 2 {\displaystyle 
  \sum_{\stackrel{{\scriptstyle \rm k}}{\rm k \neq j}}\ 
    \ft(R_{\rm jk}) } }} \quad .
\end{equation}
{The collision terms conserve energy exactly for $m^*=m$. If
however, $m^*$ is a complicated function of the momenta of all other
particles, energy conservation in the individual two-body collision
can be achieved only approximately. We find, that both definitions of
the effective mass conserve the total energy well, however,
Eq.~(\ref{eq:msr}) gives slightly better conservation and is therefore used.}

\section{Comparison with experimental data}

\begin{figure}[htbp]
\centerline{\epsfysize=60mm \epsfbox{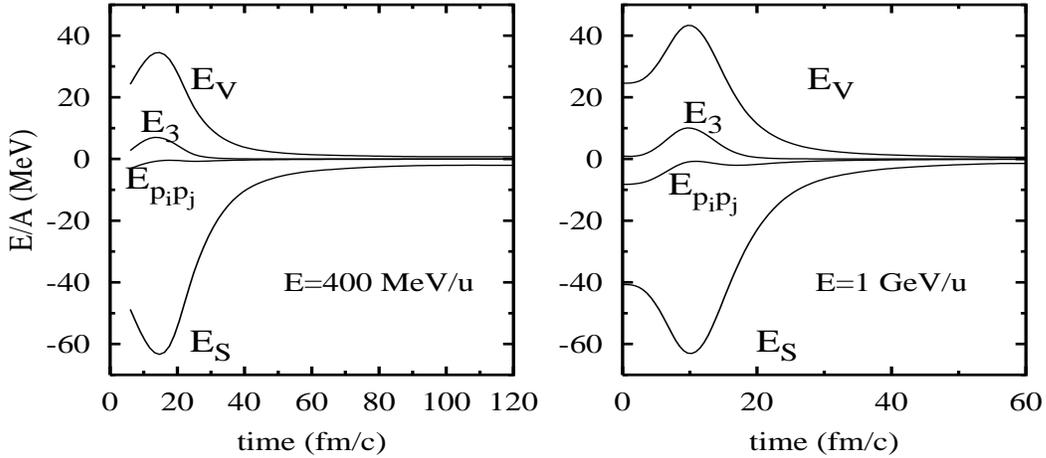}}
\caption{Time evolution of the average scalar ($E_S$) and vector
($E_V$) energy and the third order ($E_3$) and momentum dependent
($E_{p_ip_j}$) terms at 400 AMeV (left) and 1 AGeV (right). 
(See Eq.~(\protect\ref{eq:en}).)}
\label{fig-enevol}
\end{figure}
We want to apply the relativistically invariant force in
the regime of 0.5--2 AGeV bombarding energy. However, the influence
of particle creation may blur the effects of the relativistic mean-field
forces. For this reason we investigate the features of such a force at lower
bombarding energies (150--400 AMeV) and compare our results
with the available experimental data.
As a first step, we should check the approximation~(\ref{eq:en}) obtained
by expressing the four-velocities as functions of the momenta.
In Fig.~\ref{fig-enevol} the time evolution of the different energy
terms appearing in Eq.~(\ref{eq:en}) are given as the function of
time. One can see that the third order term is small and the
approximation is very good for 400 AMeV and is still acceptable
even for 1 AGeV.

In presenting our theoretical results, we generally show two kinds of
quantities: filtered and unfiltered values. The unfiltered
value is the result of the calculation. The filtered values are the
results obtained after using the experimental filter due to the
experimental hardware (reduction in the  $4\pi$ geometry, detection of
low energy particles, etc.). 
Furthermore, in both case we may apply cuts (event selection criteria),
or present the data without these cuts.

Since we wish to compare our results with the FOPI experiments, where
the most central collisions were investigated, as a first step, we examine
the connection between the experimental centrality condition and the
impact parameter in our model.

In Ref.~\cite{FOPI} it was decided that
centrality depends on the so-called ERAT values and the directivities.
These quantities were defined as
\be
\mbox{ERAT}=\frac{\sum_i p^2_{ti}/(m_i+E_i)}{\sum_i p^2_{li}/(m_i+E_i)}
\quad ,\quad D=\frac{\left| \sum_i Z_i \vec{u}_{ti}\right|}%
	{\sum_i Z_i |\vec{u}_{ti}|} \,.
\label{eq:erat}
\ee
The measured protons and nuclei have rest mass $m_i$, charge $Z_i$,
longitudinal momentum $p_{li}$ and transverse momentum $p_{ti}$. The
relation to the energy $E_i$ and the 4-velocity $u_i$ is as usual.
The sum is calculated for all {\em charged} fragments.
The (experimental) selection of the events to get 200 mb total cross
section (ERAT200) 
for 400 AMeV gold-gold collisions corresponds to ERAT$>0.66$, the 50 mb
value (ERAT50) corresponds to
ERAT$>0.88$. Experimentally, these cuts were still not found satisfactory
for selecting central events, and an additional directivity cut was applied for
the ERAT200 set with $D<0.19$ defining the ERAT200D set. The directivity
is defined in Eq.~(\ref{eq:erat}).
\begin{table}[htbp]
\begin{center}
\begin{tabular}{||c||r|r||r|r||r|r||} \hline
$b$ &\multicolumn{2}{c||}{ERAT200} & \multicolumn{2}{c||}{ERAT200D} &
	\multicolumn{2}{c||}{ERAT50} \\ \cline{2-7}
$(fm)$ & $400$  & $150$  & $400$  & $150$  
	& $400$  & $150$  \\ \hline
0.5 &  98 &  67 & 84 & 63 & 72 & 15 \\
1.0 &  95 &  64 & 48 & 51 & 36 & 14 \\
1.5 &  95 &  38 & 34 & 28 & 37 &  5 \\
2.0 &  82 &  18 & 21 &  8 &  6 &  2 \\
2.5 &  59 &  10 &  5 &  5 &  3 &  - \\
3.0 &  27 &   8 &  3 &  4 &  - &  - \\ \hline
\end{tabular}
\end{center}
\caption{Acceptance of the QMD events in percentage in the case of
different cuts for sevaral impact parameters at 400 and 150 AMeV.}
\label{tab5}
\end{table}

\begin{figure}[htbp]
\centerline{\epsfxsize=125mm \epsfbox{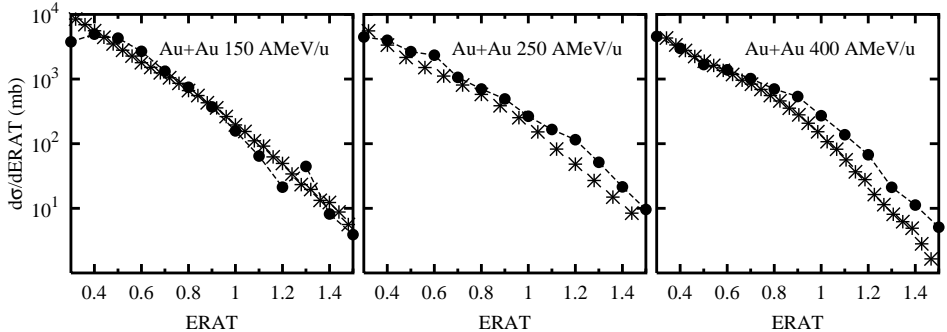}}
\caption{Measured ERAT distribution (crosses)~\cite{FOPI} at 150 AMeV (left), 
250 AMeV (middle) and 400 AMeV (right) 
and results of QMD calculations with the experimental filter
 (solid line with dots).}
\label{fig-erat}
\end{figure}

In our comparison we mostly use a cut in the impact parameter. To
see how reliable this is, we compare our total ERAT
distributions for different energies with the experimental ones. As one
can see in Fig.~\ref{fig-erat}, for 150 AMeV the agreement is excellent,
and it is still satisfactory for higher energies.

As a next step, we calculate the percentage of the accepted events for
different cuts as the function of the impact parameter.  In
Table~\ref{tab5} we show the correlation between the
impact parameter and various experimental cuts.  One can see,
that the cut ERAD200D, used mostly in the evaluation of the experimental
data, gives good centrality. In our calculation we choose the centrality
according to the impact parameter. However, to determine the importance
of the ERAT cuts, we sometimes follow the original recipe selecting
events according to their ERAT values, and compare the results of the
two selection methods. If we average over impact parameters up to
 $b=2$ fm, the values are very similar to the ones obtained with the
selection ERAT200D.

\begin{figure}[htbp]
\centerline{\epsfysize=60mm \epsfbox{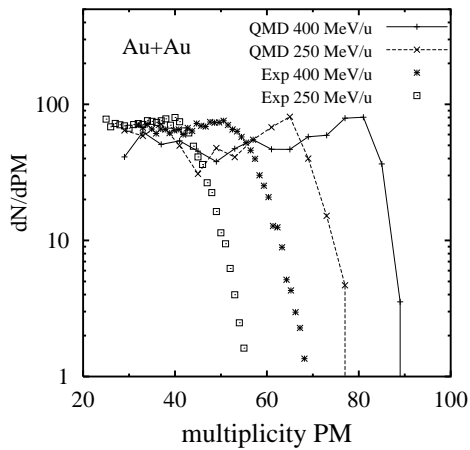}}
\caption{Measured and calculated multiplicity distribution at two beam
energies.}
\label{fig-mult}
\end{figure}
\begin{figure}[htbp]
\centerline{\epsfysize=8cm \epsfbox{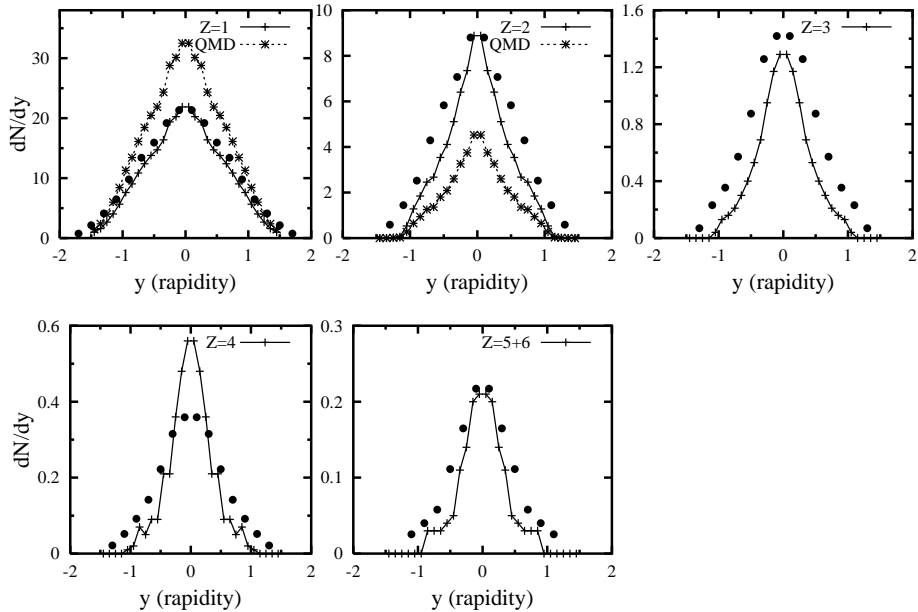}}
\caption{Measured rapidity distributions (dots)~\cite{FOPI} at 400
AMeV for various nuclear charges. The solid line with crosses is the
result of the model calculation, while the dashed lines for $Z=1$ and
$2$ with crosses correspond to the scaled model calculation (see text).}
\label{fig-rap400}
\end{figure}

Before comparing the theoretical results with the experimental ones, we
should mention a common shortcoming of all the QMD calculations.
Since shell effects are absent the number of protons is overestimated
and the abundance of alpha particles ($Z=2$) underestimated. Therefore
the total multiplicity is larger
than the experimental one (see Fig.~\ref{fig-mult}). The same phenomena
can be observed in the rapidity distributions. At 400 AMeV the
distributions of the heavier fragments are more or less agreeing with
the experimental values, however, for $Z=1$ and $Z=2$, while the
calculated shapes are good, the absolute values are not. 
If the calculated $Z=1$ distribution is scaled by a factor $2/3$ down to
the measured one and the remaining $1/3$ of the protons is appointed to
the $Z=2$ distribution a consistent picture is achieved,
as is shown in Fig.~\ref{fig-rap400}.

\begin{figure}[htbp]
\centerline{\epsfysize=60mm \epsfbox{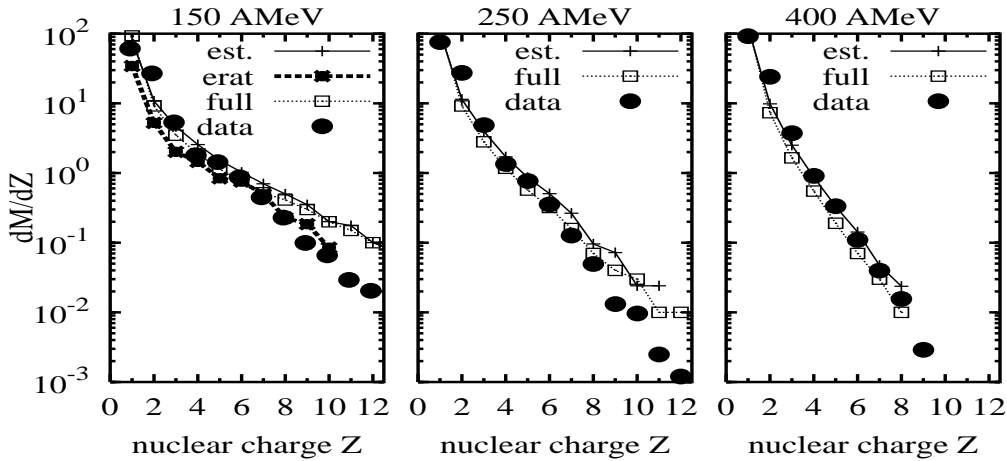}}
\caption{Measured (dots) and calculated charge distributions at three
incident energies. The solid line with crosses corresponds to filtered
QMD events for $b\le 1.5$fm extrapolated to $4\pi$ geometry (est.), while
the dashed line with squares is the QMD unfiltered calculation 
for $b\le 1.5$fm. For 150 AMeV the calculation performed with ERAT200D
cut (thick dashed line with crosses) is also presented.}
\label{fig-charge}
\end{figure}

The measured (4$\pi$ averaged) charge distributions also agree well with
the calculated ones. The tail of the distribution obtained with an
impact parameter cut $b\le 1.5$fm in the calculations at 150 AMeV slightly
differs from the experimental one, however, when applying the ERAT200D cut
on the calculated events, the overestimation for large charge numbers
disappear (see Fig.~\ref{fig-charge}). This means that the ERAT200D cut
throws out those central events which have bigger fragments.
For higher energies
the none of the ERAT cuts modify the picture significantly which is in
accord with the experimental findings~\cite{FOPI}.

\begin{figure}[htbp]
\centerline{\epsfysize=100mm \epsfbox{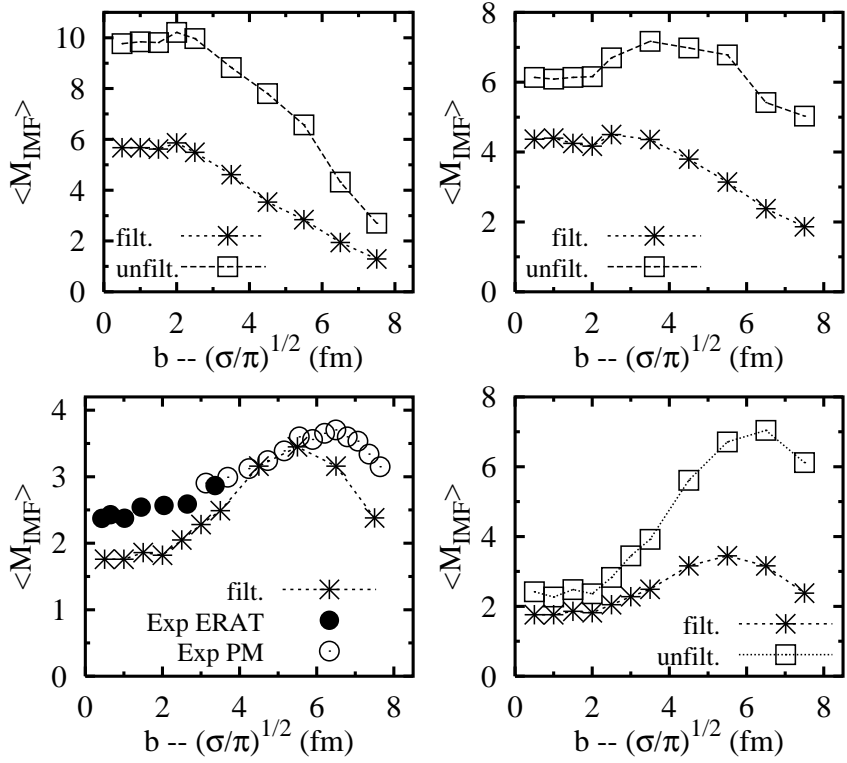}}
\caption{Number of IMF-s as a function of impact parameter
at three incident energies (150 AMeV -- upper left, 250 AMeV -- upper
right and 400 AMeV -- lower line) with and without filtering the QMD
calculation, and the experimental values at 400 AMeV~\cite{FOPI}.}
\label{fig-imfb}
\end{figure}

One important quantity which is characteristic of multifragmentation is
the number os intermediate mass fragments (IMF, $2<Z<30$).
The number of IMFs is a crucial test for the capability of the model to
describe many-body correlations. There is however the complication that
the fragments are altered by subsequent decay before they reach the
counter. The modification of the distributions will be stronger if the
fragments are created with high excitation energy.
At the end of our QMD calculation the fragments are bound, with an
excitation energy of 1--2 MeV per particle. At these energies
deexcitation occurs mainly through radiation and
sometimes by particle emission, therefore the final mass numbers of
the fragments will change only slightly.  In Fig.~\ref{fig-imfb}, we show
the filtered and unfiltered IMF distributions as the function of the
impact parameter for the different energies. 
Contrary to most theoretical
calculations, which substantially underestimate the IMF distribution in
central collisions, our model yields values much closer to the measured
ones.  
We also studied how the number of IMFs depend on the assumption or
approximations made in the model.

To determine the effect of the effective mass on multifragmentation, we
used for $m^*$ both of the expressions~(\ref{eq:msr}) and (\ref{eq:msi})
and in addition we also
performed calculations with $m^*$ set to m~\cite{Judit} in the collision
term.
  
The choice of the effective mass influenced the number of IMF values
less than 6\%. The choice $m^*=m$ in the collision term always yielded
the largest $\langle M_{IMF}\rangle$.
Neglecting the third order term in Eq.~(\ref{eq:en}), but readjusting
the coupling strengths such that the saturation density was kept at its
correct value, the change was
less than 3\%, thus completely negligible. A 20\% change in the
collision cross section also gave less than 7\% change. 

\begin{figure}[htbp]
\centerline{\epsfysize=5cm \epsfbox{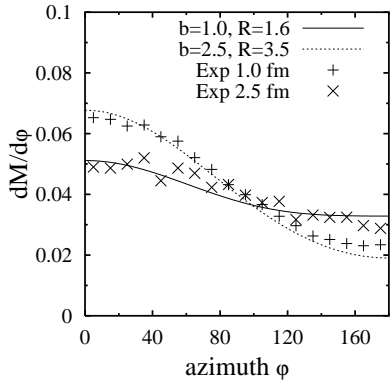}}
\caption{Azimuthal angle distribution of intermediate mass fragments
using the ERAT filter for 250 AMeV. Dots represent the
experimental data~\cite{FOPI}, while the solid lines are
fits~(\protect\ref{eq:phifit}) to the QMD calculations.}
\label{fig-phi}
\end{figure}

In addition we calculated the azimuthal distribution of the IMFs at 250
AMeV, and found a good agreement with the experimental values.
In Fig.~\ref{fig-phi} the experimental data~\cite{FOPI} are compared to
\be
  dM/d\phi&=&a_0 (1\!+\!a_1\cos\phi\!+\!a_2\cos{2\phi})
\label{eq:phifit}
\ee
which was fitted to the calculated numbers. The ratio 
 $R=\frac{dM/d\phi(\phi\!=\!0)}{dM/d\phi(\phi\!=\!\pi)}$ is also indicated.

\begin{figure}[htbp]
\centerline{\epsfysize=8cm \epsfbox{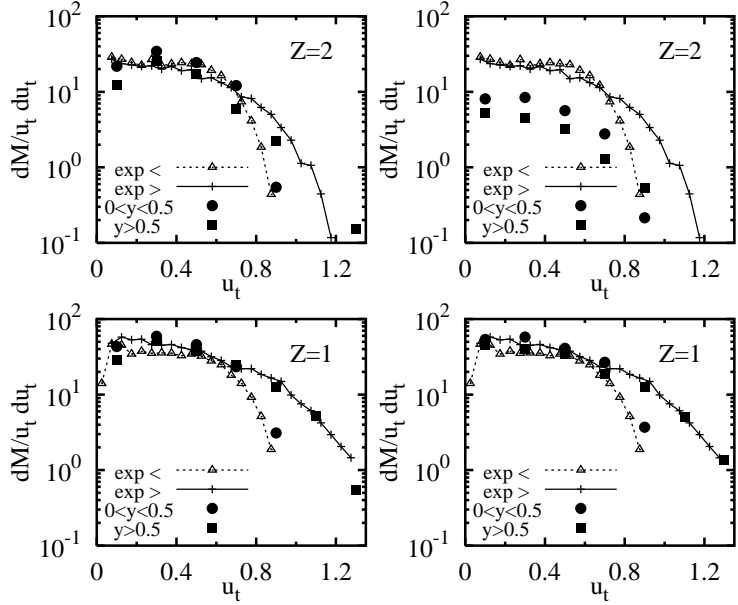}}
\caption{Invariant transverse four-velocity spectra at 150 AMeV with
ERAT200D cuts (left block) and at $0<b\le 2.5 {\rm fm}$ (right block)
for protons and alpha particles.}
\label{fig-ut150}
\end{figure}

\begin{figure}[htbp]
\centerline{\epsfysize=40mm \epsfbox{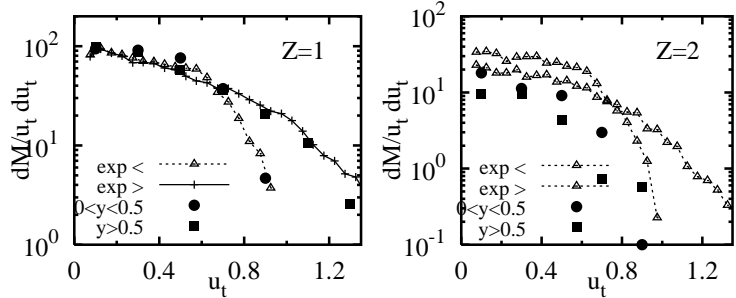}}
\caption{Invariant transverse four-velocity spectra at 400 AMeV and
 $0<b\le 2 {\rm fm}$ for $Z=1$ and $2$.}
\label{fig-ut400}
\end{figure}

In Figs.~\ref{fig-ut400} and \ref{fig-ut150} the scaled transverse
four-velocity distribution $dM/u_tdu_t$ is shown
for forward particles (with $y>0.5$) and for particles around
midrapidity ($y<0.5$). The lines are taken from the experimental
data~\cite{FOPI}, while the dots and squares represent 
the results of the QMD calculations for central collisions in the impact
parameter range $0<b\le 2.5 {\rm fm}$ and the ERAT200D cut.
For $Z=1$ and $E=150 AMeV$ the ERAT200D cut has little influence. Both
calculated spectra agree with the data. For $Z=2$ the situation is
different, the ERAT200D cut seems to select events which have about 5
times more $\alpha$ particles.

Fig~\ref{fig-ut150} shows agreement inthe spectra for $E=400 AMeV$ and
 $Z=1$ while the calculated $Z=2$ spectra deviate substantially. Here
 the ERAT200D selection has little influence on the spectra in contrast
 to the $E=150 AMeV$ displayed in Fig.~\ref{fig-ut400}.
For 400 AMeV energy there is no
deviation between the ERAT cut and the impact parameter cut.
 
\section{Collective flow and the kinetic energies of large fragments}

\begin{figure}[htbp]
\centerline{\epsfysize=5cm \epsfbox{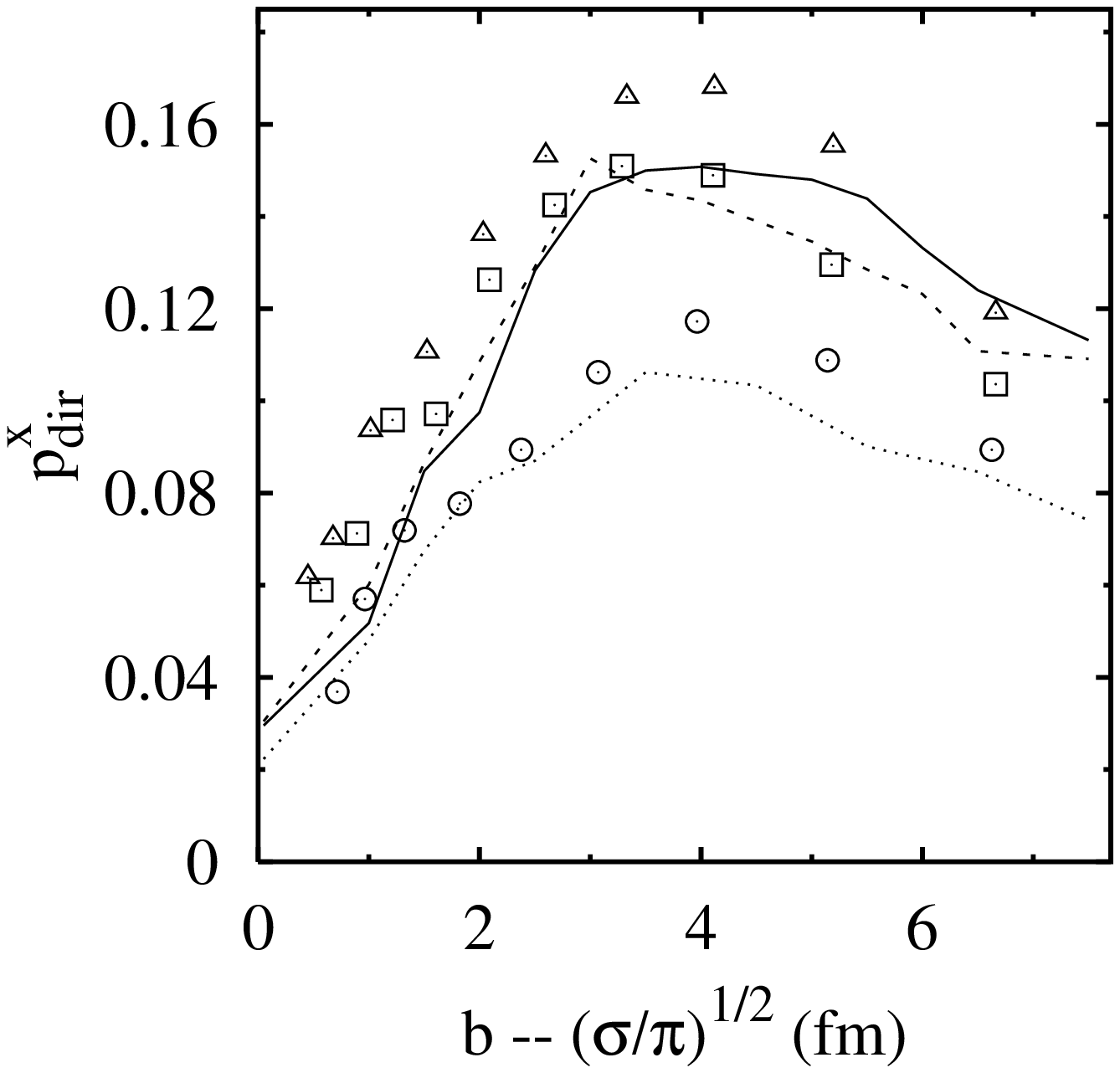} \hfil
\epsfysize=5cm \epsfbox{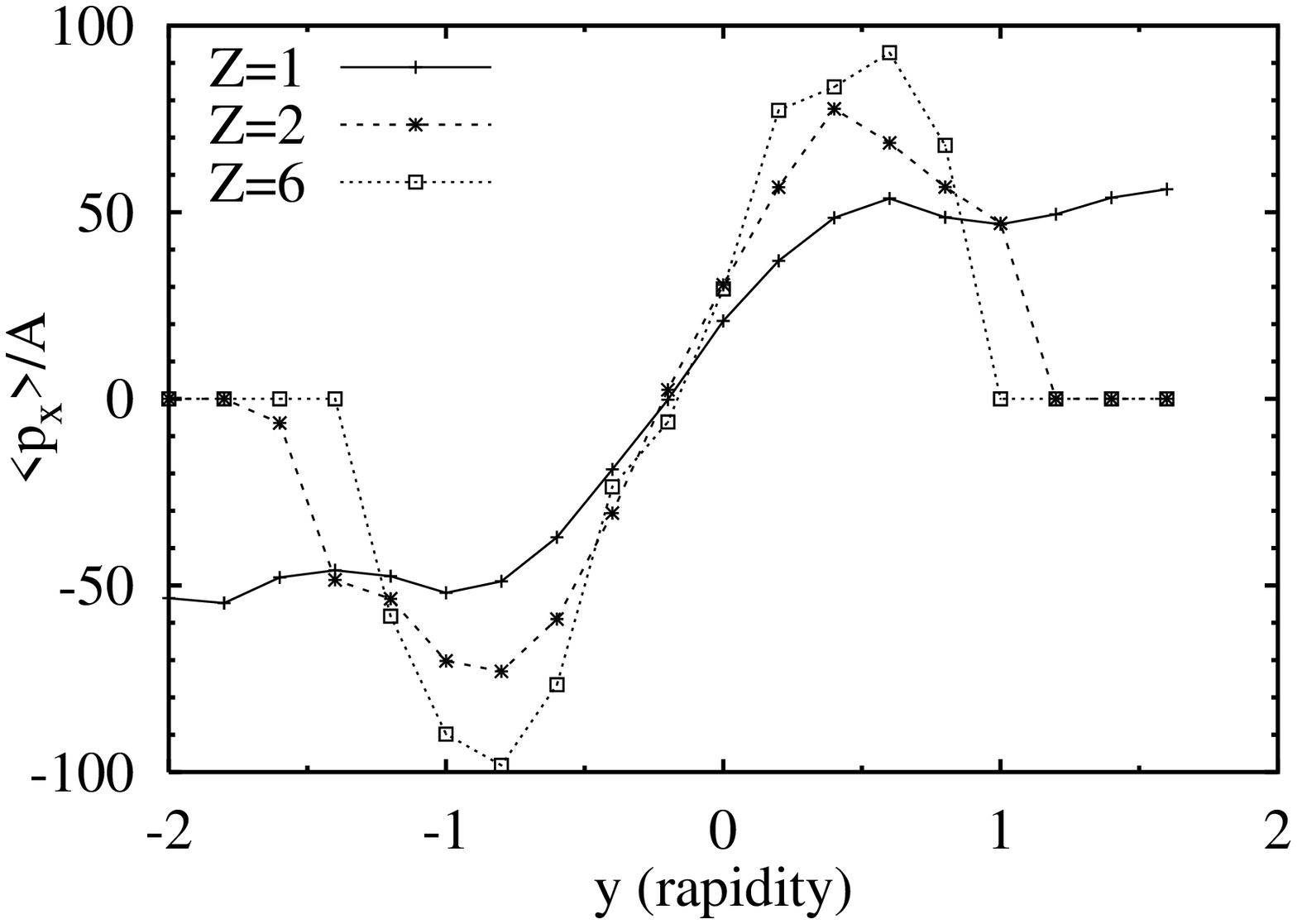}}
\caption{{\bf a} Measured (symbols) and calculated (lines) integrated
scaled side flow as the function of the impact parameter at an incident
energy of 150 (triangle and solid line), 250 (square and dashed line) 
and 400 AMeV (circle and dotted line) (left).  {\bf b} Side flow for
various nuclear 
charges at an incident energy of 400 AMeV as the function of the
rapidity (right).}
\label{fig-pxdir}
\end{figure}

Due to its sensitivity, flow is one of the most important data provided
by the experiments to test the theoretical models. One can distinguish
three types of flow: the side, the squeeze-out and the spherical. The
first two are well known, and have been studied extensively both
experimentally~\cite{GUS} and theoretically~\cite{PAWEL} for a
long time. We present the integrated side flow,
\be
	p_x^{\mbox{dir}} = \sum_i Z_i u_{xi} / \sum_i Z_i \,,
\ee
in Fig.~\ref{fig-pxdir}a as the function of the impact parameter. The
agreement with the experimental values is remarkable.

In Fig.~\ref{fig-pxdir}b we present the side flow for different fragment
masses as the function of the rapidity at bombarding energy of 400
AMeV. The side flow increases with the increase of the fragment mass,
which is in accordance with the experimental observation~\cite{GUS}.

\begin{figure}[htbp]
\centerline{\epsfysize=5cm \epsfbox{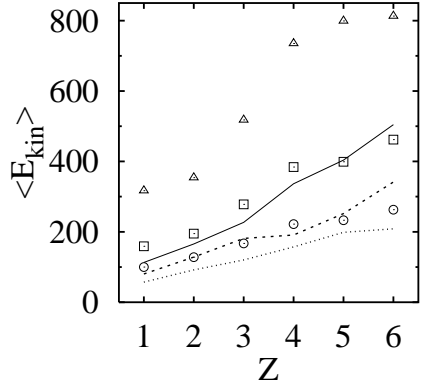}}
\caption{Measured (symbols) and calculated (lines) average kinetic
energy as the function of the fragment charge number for 150 (circles
and dotted line), 250 (squares and dashed line) and 400 AMeV (triangles
and solid line),
summed up to $b=2.5 fm$ impact parameters.}
\label{fig-mkin}
\end{figure}

Recently interest has been focused on a third type of flow both
experimentally~\cite{FOPI} and theoretically~\cite{My2}, namely the so
called central (or radial) flow, which is most prominent in central
collisions where the side flow already disappears.  Experimentally, it
was found, that the kinetic energy of the fragments increases as their
mass number increases.  Although our calculation does not reproduce the
experimental amount of the kinetic energy, we also obtained a curve
which shows a significant increase (see Fig.~\ref{fig-mkin}).

\begin{figure}[htbp]
\centerline{\epsfysize=40mm \epsfbox{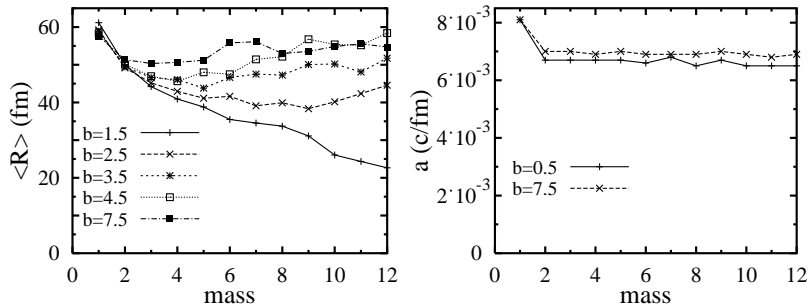}}
\caption{{\bf a.} Average distance of the fragments (in $fm$) from the
center of mass as the function of their mass at 400 AMeV (left). 
{\bf b.} The radial velocity parameter, $a^{(M)}$, as the
function of the fragment mass at 400 AMeV (right). Both curves
were calculated at t=150 fm/c.}
\label{fig-dist}
\end{figure}

The reason for the difference between the theoretical and experimental
values is the following: on examining the structure of the $A>6$
fragments, it can be observed that they are sometimes a mixture of
target and projectile nucleons. Since this means that the nucleons in
the fragments collided many times, at the freeze-out density they are
still not too far from the center, hence their kinetic energy is
small. It also can occur that the fragments consist almost completely of
either target, or projetile nucleons, which move together during the
collision and only rarely interact with other nucleons.  In this case
the kinetic energy of the fragment is large. However, these fragments
move mainly in a forward direction, so their ERAT value is smaller than
the experimental cut will allow. In Fig.~\ref{fig-dist}a we give the
fragments' average distance from the center as the function of their
mass, where the above-mentioned tendency is clearly seen.  For large
impact parameters, the large and small fragments move together.

If we determining the velocity of the fragment from an assumed blast
profile,
\be
  \vec{u}_{\rm i}=a\vec{r}_{\rm i} \quad ,\quad
	a^{(M)}=\frac{\sum_{\rm i} \vec{u}_{\rm i} \vec{r}_{\rm i}}%
		{\sum_{\rm i} r_{\rm i}^2}
\ee
and perform the summation only for clusters with mass $M$, we can also
obtain the $a$ value as the function of the mass.  Fig.~\ref{fig-dist}b
show, that $a$ is almost independent of the mass.

Experimentally, the central flow energy is evaluated with the help of
the blast model from the mass number dependence of the measured kinetic
energies. In this method, the contribution of the small number of the
larger fragments dominates the final result. Since in our case the
kinetic energies of the big fragments are not large enough, we had to
apply a different method to determine the central flow.

Considering that the total mass of the large fragments together is small
compared to the total mass of the two colliding nuclei, we decided to
determine the central flow and flow energy using all the
particles~\cite{flow}. We divided the total solid angle into 32 parts
and calculated the average velocity, as much as the number of particles
in each section. The fluctuation of the averages was less than 2\% and
4\%, respectively at 400 AMeV for central events ($b\le 1 fm$).

Our conclusion was that up to an impact parameter of 2 fm there is a
significant central flow, and the concept of thermal energy, which is
based on the thermalization, can be used. The values calculated from our
model agree with the ones which were obtained from the experiments using
the blast model (see Table~\ref{tab-vel} and for the detailes
Ref.~\cite{flow}).
\renewcommand\baselinestretch{0.9}
\begin{table}
\caption{Flow energies (MeV), ``thermal'' energies and average flow 
velocities ($\beta$) at 400
(upper), 250 (middle) and 150 AMeV (lower part) at an impact parameter
of b=1 fm.
}
\label{tab-vel}
\begin{center}
\begin{tabular}{||r||c|c||c|c||c|c||}\hline\hline
 A MeV & E$_{\rm flow}$ & E$^{exp}_{\rm flow}$
	& E$_{th}$ & E$^{exp}_{th}$ & 
	$\beta$ & $\beta^{exp}$ \\ \hline
400& 59.7 & 57.8 & 32.7 & 32.8 & 0.338 & 0.334
	\\ 
250& 34.8 & 34.0 & 16.8 & 21.5 & 0.264 & 0.263
	\\ 
150& 18.3 & 19.9 & 10.2 & 12.6 & 0.194 & 0.204 \\ \hline\hline
\end{tabular}
\end{center}
\end{table}
\renewcommand\baselinestretch{1}

\section{Freeze-out}

An important concept for understanding the dynamics of
multifragmentation is the occurance of freeze-out. After the violent
initial phase of the collision with a rapid increase of entropy, density
and local excitation energy the system begins to expand and cool. Freeze
out characterizes the situation where the particles are far enough from
each other so that they are not colliding any more. If this happens
throughout space in a narrow time interval one may define a freeze-out
time.
\begin{figure}[htbp]
\centerline{\epsfysize=45mm \epsfbox{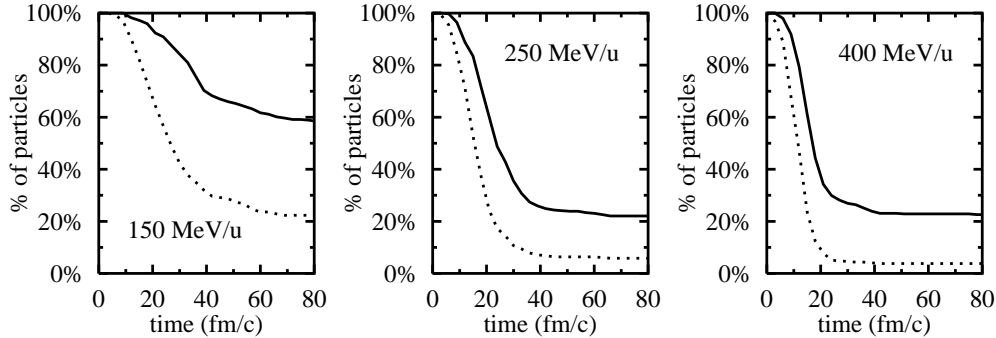}}
\caption{Number of particles which collided less than three times (solid
line), and which did not collide at all (dotted line) for three incident
energies, at an impact parameter of $b=0.5 fm$, as the function of
time.}
\label{fig-cold}
\end{figure}

Depending on the speed of the expansion freeze-out will ``freeze'' the
mass distribution and the velocity of the nucleons and cluster
in different situations on the way towards equilibration.
One way of recognizing that freeze-out is happening is to observe that
the number of unblocked collisions are drastically decreasing. It is clear
that some collision can still occur even after freeze out, because some
of the nucleons which lost their energy and are still near the center
(about 5\% of all nucleons) are nevertheless close enough to each other to be
able to collide.  The nucleons within the clusters can also collide with
each other, however, the number of such collisions is small. In
Fig.~\ref{fig-cold}, we show  as the function of time the fraction of
nucleons which did not
collide at all and those that collided less than three times.
For 400 AMeV and 200 AMeV one observes that already after 30 fm/c the
fraction of particles which did not suffer a collision does not decrease
anymore because they have escaped. The same holds true for those which
collide three or more times (percentage above full line) so that one can
assume that after 30 fm/c no further equilibration takes place.

\begin{figure}[htbp]
\centerline{\epsfysize=4cm \epsfbox{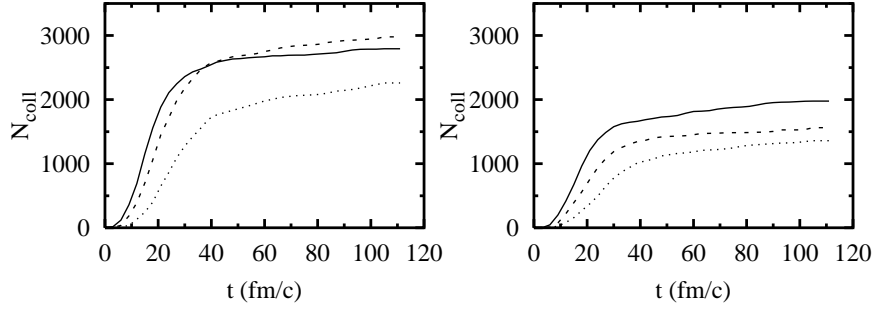}}
\caption{Time evolution of the number of collisions
at an impact parameter of b=0.5 fm (left) and b=5.5 fm (right)
for 150 AMeV (dotted), 250 AMeV (dashed) and 400 AMeV (solid line).}
\label{fig-timec}
\end{figure}

In Fig.~\ref{fig-timec}, the number of collisions is displayed as the
function of time for 150, 250 and 400 MeV. Also From these results it
can be seen, that freeze-out occurs around 30-40 fm/c. From
Fig.~\ref{fig-enevol} it can be noted that this is also the time at
which the average mean field energy drops drastically.
For lower energies, the transition from the interacting to the freezed
out system is not so sharp. However, even there one can determine that the
average freeze-out time is around 40 fm/c~\cite{My1}.
\begin{figure}[htbp]
\centerline{\epsfysize=8cm \epsfbox{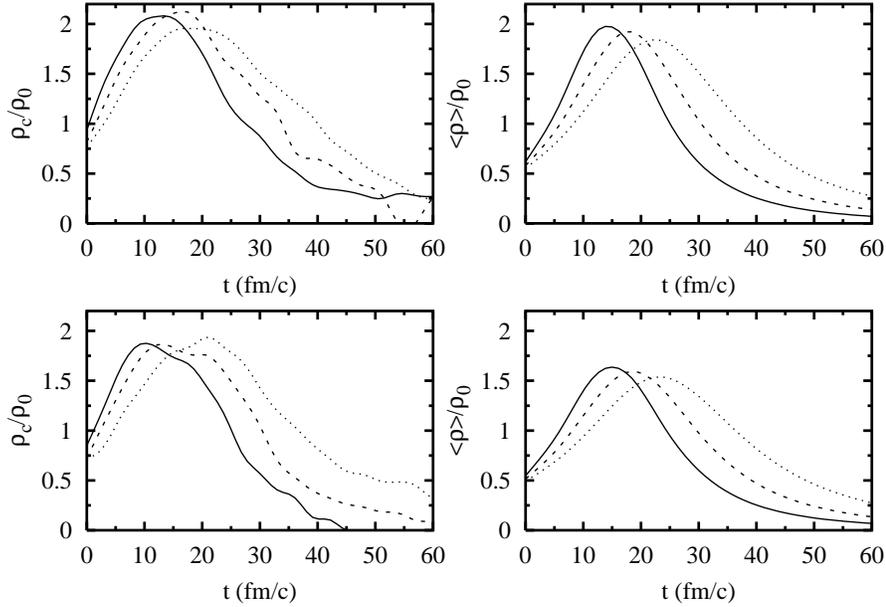}}
\caption{Time evolution of the central (left) and average (right)
density at an impact parameter of b=0.5 fm (upper) and b=5.5 fm (lower)
for 150 AMeV (dotted), 250 AMeV (dashed) and 400 AMeV (solid line).}
\label{fig-timed}
\end{figure}

\begin{figure}[htbp]
\centerline{\epsfysize=4cm \epsfbox{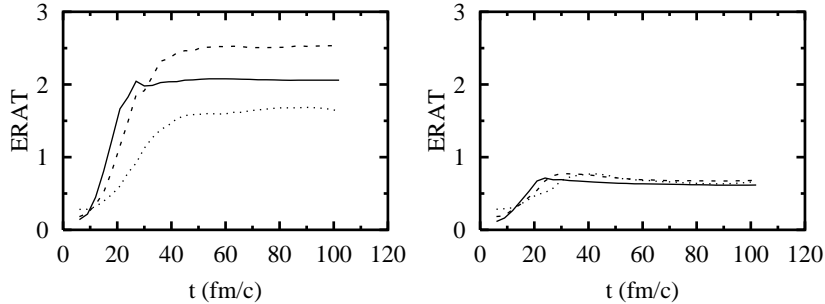}}
\caption{Time evolution of the ERAT value
at an impact parameter of b=0.5 fm (left) and b=5.5 fm (right)
for 150 AMeV (dotted), 250 AMeV (dashed) and 400 AMeV (solid line).}
\label{fig-timee}
\end{figure}

In Fig.~\ref{fig-timed}, we show the average and central densities of
the colliding system as a function of time. The central density was
determined in a radius of 2 fm around the center of mass, while for the
average density the r.m.s radius of the colliding system was used.  In
Fig.~\ref{fig-timee}, the ERAT values are given as the function of time.
All of these figures confirm the above given freeze-out time of
approx. 40 fm/c. We find the average density at the freeze-out time to
be $0.3\varrho_0$ for 400 AMeV and around $0.5\varrho_0$ at 150 AMeV, in
agreement with other model predictions~\cite{My2,My1}


\section{Conclusion}

In this paper we derived a relativistically covariant Hamiltonian and
the corresponding equations of motion starting from a Walecka type
lagrangian with
derivative coupling (Zim\'anyi-Moszkowski lagrangian).  
{ The relativistic covariant lagrangian determines the momentum
dependence of the forces in the mean field approximation
uniquely. Despite the fact that QMD is not a completely relativistic
model, the derived momentum dependence for momenta small compared to the
rest mass reflects the original relativistically covariant theory.}
The equations of motion 
can be solved easily in a QMD code for the
approximation we introduced. 
We applied these equations to gold on
gold collisions at 150, 250 and 400 AMeV, and compared our results with
those of the FOPI experiment~\cite{FOPI}.

We found that the experimental filtering criteria (ERAT200D) selects the
very central events for 400 AMeV, and approximately central events for
lower energies. The results obtained from our model are in agreement
with most of the experimental observatories, however, the mean kinetic
energies of large fragments are smaller than the measured ones.

The calculated IMF numbers are found to be close to the experimental
results, but still lower. Furthermore, our calculations
show that these numbers are rather insensitive to most details of the
{ equation of state},
the only relevant quantity is the saturation density.

The model reproduces most of the experimental flow results.
The fact that the mean central flow velocities agree with the
experimental ones 
for all three beam energies indicates, that 
the momentum dependence of the applied mean-field force { might be}
satisfactory.
We would like to emphasize again that this momentum dependence
is a result of the relativistic lagrangian and contains no additional
parameters.

The freeze-out time for central collisions is between 30 and 50 fm/c,
depending on the beam energy. The transition to the frozen stage is more
pronounced for higher energies. The freeze-out density for central
collisions is around third--half the nuclear matter saturation density
depending on the beam energy, and smaller for higher energies.

We believe that the Zim\'anyi-Moszkowski lagrangian, which assumes
a derivative coupling between the scalar field and the nucleons,
provides an adequate momentum dependent mean-field not only for nuclear
matter where the adjustment of the coupling strengths
to saturation density and binding energy provides also the proper
momentum dependence { of the optical potential~\cite{Lindner}}, 
but also for heavy ion collisions where the proper momentum dependence
is essential.

The energy is well conserved during the collision process: even in
the case of 400 AMeV beam energy, the maximal deviation
is less than 0.8 MeV per particle. This shows that the individual
nucleon-nucleon collision process and the effective mass is reasonably
treated.

We found that the derived relativistically covariant Hamiltonian
gives good results with the QMD code in the energy range 
of 150--400 AMeV. Since the approximations applied are still valid at
1 AGeV this model seems to be appropriate to study higher 
energy collisions, where particle production starts to be important.
This work is in progress.

\section*{Acknowledgments}

One of the authors (J.N.) should like to express her thanks to
Prof. W. Greiner and the University of Frankfurt and to
Prof. N\"orenberg and the GSI for their kind hospitality, during her
visit to this institutions, where part of 
this work was done. Discussions with W. Reisdorf and A. Gobbi
are highly acknowledged. We express our thanks to the FOPI group
providing the experimental filter program.
This work was partly supported by 
Hungarian OTKA grant T022931 and 
FKFP grant 0126/1997.


\begin{thebibliography}{99}

\bibitem{FOPI}
W. Reisdorf et al.,
  Nucl. Phys. {\bf A612} (1997) 493.

\bibitem{USHIP}
H. Feldmeier, J. N\'emeth and G. Papp,
  APH Heavy Ion Physics {\bf 3} (1996) 71.

\bibitem{experiment} 
FOPI collaboration,
	Nucl. Phys. {\bf A586} (1995) 755;
M. Petrovici et al.,
	Phys. Rev. Lett. {\bf 74} (1995) 5001;
S.C. Jeong et al.,
        Phys. Rev. Lett. {\bf 72} (1994) 3468;
W.C. Hsi et al.,
        Phys. Rev. Lett. {\bf 73} (1994) 3367.
 
\bibitem{BUU}
G.F. Bertsch and S. Das Gupta,
	Phys. Rep. {\bf 160} (1988) 189;
W. Cassings, V. Metag, U. Mosel and K. Niita,
	Phys. Rep. {\bf 188} (1990) 363.

\bibitem{AICH} 
J. Aichelin, 
	Prog. Nucl. Part. Phys. {\bf 30} (1991) 191;
J. Aichelin, G. Peilert, A. Bohnet, A. Rosenhauer, H. St\"ocker and W. Greiner,
	Phys. Rev. {\bf C37} (1988) 2451;
J. Aichelin, 
	Phys. Rep. {\bf 202} (1991) 233.

\bibitem{frank} 
G. Peilert, H. St\"ocker and W. Greiner, 
	Rep. Prog. Phys. {\bf 57} (1994) 533;
J. Konopka, 
	Prog. Part. and Nucl. Phys. {\bf 30} (1993) 301;
C. Hartnack {\it et al.}, 
	Nucl. Phys. {\bf A580} (1994) 643.

\bibitem{boguta}
J. Boguta and A.R. Bodmer,
	Nucl. Phys. {\bf A292} (1977) 413.

\bibitem{zimmos} 
J. Zim\'anyi and S.A. Moszkowski, 
	Phys. Rev. {\bf C42} (1990) 1416

\bibitem{Fel2}
H. Feldmeier, 
	in ``{\it Relativity in General}'', eds. J. Diaz Alonso and M. Lorente
	Paramo, (Editions Fronti\`eres 1994, ISBN 2-86332-168-4).

\bibitem{Wheeler}
J.A. Wheeler and R.P. Feynman,
	 Rev. Mod. Phys. {\bf 21} (1949) 425.

\bibitem{amd}
A. Ono, H. Horiuchi, T. Maruyama and A. Ohnishi,
	Phys. Rev. Lett. {\bf 68} (1992) 2898;
	Prog. Theor. Phys. {\bf 87} (1992) 1185.

\bibitem{fmd} 
H. Feldmeier;
  Nucl. Phys. {\bf A515} (1990) 147;
H. Feldmeier, K. Bieler and J. Schnack,
  Nucl. Phys. {\bf A586} (1995) 493.

\bibitem{Lindner}
H. Feldmeier and J. Lindner,
	Z. Physik {\bf A341} (1991) 83.

\bibitem{BelMartin}
L. Bel and J. Martin, 
	Ann. Inst. Henri Poincar\'e, A XXII (1975) 173;
L. Bel and J. Martin, 
	Phys. Rev. {\bf D9} (1974) 2760.

\bibitem{Jackson}
J.D. Jackson, 
	Classical Electrodynamics, (New York: John Wiley \& Sons, Inc., 1962).

\bibitem{serwal} 
B.D. Serot and J.D. Walecka,
	Adv. Nucl. Phys. {\bf 16} (1986) 1.

\bibitem{judqmd}
J. N\'emeth, C. Ng\^o, H. Ng\^o and L. De Paula,
	XX. International Workshop in Hirschegg (1992) p262.

\bibitem{paula}
L. De Paula, J. N\'emeth, Sa Ben Hao, S. Leray, C. Ng\^o, 
S.R. Souza and Zheng Yu Heng, 
	Phys. Lett. {\bf B258} (1991) 251;
S.R. Souza, L. De Paula, S. Leray, J. N\'emeth, C. Ng\^o and H. Ng\^o,
	Nucl. Phys. {\bf A571} (1994) 159.

\bibitem{cug} 
J. Cugnon, T. Mizutani and J. Vandermeulen, 
	Nucl. Phys. {\bf A352} (1981) 505.

\bibitem{wolf} 
Gy. Wolf, G. Batko, W. Cassing, U. Mosel, K. Niita and M. Sch\"afer, 
	Nucl. Phys. {\bf A517} (1990) 615.

\bibitem{Sitg93}
J. N\'emeth and G. Papp,
  in {\it ``Dynamical features of nuclei''}, p36; 
	ed. by X. Vinas, World Scientific, 1994.

\bibitem{Judit}
J. N\'emeth,
	Heavy Ion Physics {\bf 5} (1997) 249.

\bibitem{GUS}
H. A. Gustafsson et al.,
        Phys. Rev. Lett. {\bf 52} (1984) 1590;
H. H. Gutbrod, A. M. Poskanzer and H. G. Ritter,
        Rep. Prog. Phys. {\bf 52} (1989) 1267.
 
\bibitem{PAWEL}
P. Danielewicz,
	Phys. Rev. {\bf C51} (1995) 716;
J. Konopka, H. St\"ocker and W. Greiner,
	Proc. XXII. Hirschegg Workshop, Hirschegg, Austria 1994, p218.

\bibitem{My2}
W. Reisdorf,
        XXII. Hirschegg Workshop (1994) p93;
W. N\"orenberg and G. Papp,
	{\it in} ``Critical Phenomena and Collective Observables'',
ed. S. Costa, S. Albergo, A. Insola, C. Tuv\'e; World Scientific, 1996,
Singapore, p. 377.

\bibitem{flow}
J. N\'emeth and G. Papp,
	nucl-th/9711039, to appear in Phys. Rev. {\bf C}.

\bibitem{My1}
G. Papp and W. N\"orenberg,
	Heavy Ion Physics {\bf 1} (1995) 241;
G. Papp, J. N\'emeth and J.P. Bondorf, 
	Phys. Lett. {\bf B278} (1992) 7.

\end{thebibliography}
\end{document}